\pdfoutput=1
\newcommand\apjcls{1}
\newcommand\aastexcls{2}
\newcommand\othercls{3}

\newcommand\papercls{\aastexcls}
\documentclass[tighten, times, trackchanges, preprint2]{aastex6}

\if\papercls \apjcls
\usepackage{apjfonts}
\else\if\papercls \othercls
\usepackage{epsfig}
\fi\fi
\usepackage{ifthen}
\usepackage{natbib}
\usepackage{amssymb, amsmath}
\usepackage{mathrsfs}
\usepackage{bm}
\usepackage{appendix}
\usepackage{etoolbox}
\usepackage[T1]{fontenc}
\usepackage{paralist}

\if\papercls \apjcls
\newcommand\aas{\ref@jnl{AAS Meeting Abstracts}}
\newcommand\dps{\ref@jnl{AAS/DPS Meeting Abstracts}}
\newcommand\maps{\ref@jnl{MAPS}}
\else\if\papercls \othercls
\usepackage{astjnlabbrev-jh}
\fi\fi

\if\papercls \aastexcls
\hypersetup{citecolor=blue, 
	linkcolor=blue, 
	menucolor=blue, 
	urlcolor=blue}  
\else
\usepackage[
bookmarks=true,           
bookmarksnumbered=true,   
colorlinks=true,          
citecolor=blue,           
linkcolor=blue,           
menucolor=blue,           
urlcolor=blue,            
linkbordercolor={0 0 1},  
pdfborder={0 0 1},
frenchlinks=true]{hyperref}
\fi

\if\papercls \othercls

\else

\fi

\providecommand{\adsurl}[1]{\href{#1}{ADS}}

\makeatletter
\patchcmd{\NAT@citex}
{\@citea\NAT@hyper@{%
		\NAT@nmfmt{\NAT@nm}%
		\hyper@natlinkbreak{\NAT@aysep\NAT@spacechar}{\@citeb\@extra@b@citeb}%
		\NAT@date}}
{\@citea\NAT@nmfmt{\NAT@nm}%
	\NAT@aysep\NAT@spacechar\NAT@hyper@{\NAT@date}}{}{}

\patchcmd{\NAT@citex}
{\@citea\NAT@hyper@{%
		\NAT@nmfmt{\NAT@nm}%
		\hyper@natlinkbreak{\NAT@spacechar\NAT@@open\if*#1*\else#1\NAT@spacechar\fi}%
		{\@citeb\@extra@b@citeb}%
		\NAT@date}}
{\@citea\NAT@nmfmt{\NAT@nm}%
	\NAT@spacechar\NAT@@open\if*#1*\else#1\NAT@spacechar\fi\NAT@hyper@{\NAT@date}}
{}{}
\makeatother

\makeatletter
\DeclareRobustCommand{\lowcase}[1]{\@lowcase#1\@nil}
\def\@lowcase#1\@nil{\if\relax#1\relax\else\MakeLowercase{#1}\fi}
\makeatother

\DeclareSymbolFont{UPM}{U}{eur}{m}{n}
\DeclareMathSymbol{\umu}{0}{UPM}{"16}
\let\oldumu=\umu
\renewcommand\umu{\ifmmode\oldumu\else\math{\oldumu}\fi}

\if\papercls \othercls

\else

\fi

\let\oldsim=\sim
\renewcommand\sim{\ifmmode\oldsim\else\math{\oldsim}\fi}
\let\oldpm=\pm
\renewcommand\pm{\ifmmode\oldpm\else\math{\oldpm}\fi}
\newcommand\by{\ifmmode\times\else\math{\times}\fi}

\newcommand\tablebox[1]{\begin{tabular}[t]{@{}l@{}}#1\end{tabular}}
\newbox{\wdbox}
\renewcommand\c{\setbox\wdbox=\hbox{,}\hspace{\wd\wdbox}}
\renewcommand\i{\setbox\wdbox=\hbox{i}\hspace{\wd\wdbox}}

\newcount\timect
\newcount\hourct
\newcount\minct
\newcommand\now{\timect=\time \divide\timect by 60
	\hourct=\timect \multiply\hourct by 60
	\minct=\time \advance\minct by -\hourct
	\number\timect:\ifnum \minct < 10 0\fi\number\minct}

\catcode`@=11

\newcommand\comment[1]{}

\newcommand\commenton{\catcode`\%=14}

\renewcommand\math[1]{$#1$}
\newcommand\mathshifton{\catcode`\$=3}

\let\atab=&
\newcommand\atabon{\catcode`\&=4}

\let\oldmsp=\sp
\let\oldmsb=\sb
\def\sp#1{\ifmmode
	\oldmsp{#1}%
	\else\strut\raise.85ex\hbox{\scriptsize #1}\fi}
\def\sb#1{\ifmmode
	\oldmsb{#1}%
	\else\strut\raise-.54ex\hbox{\scriptsize #1}\fi}
\newbox\@sp
\newbox\@sb
\def\sbp#1#2{\ifmmode%
	\oldmsb{#1}\oldmsp{#2}%
	\else
	\setbox\@sb=\hbox{\sb{#1}}%
	\setbox\@sp=\hbox{\sp{#2}}%
	\rlap{\copy\@sb}\copy\@sp
	\ifdim \wd\@sb >\wd\@sp
	\hskip -\wd\@sp \hskip \wd\@sb
	\fi
	\fi}
\def\msp#1{\ifmmode
	\oldmsp{#1}
	\else \math{\oldmsp{#1}}\fi}
\def\msb#1{\ifmmode
	\oldmsb{#1}
	\else \math{\oldmsb{#1}}\fi}

\def\supon{\catcode`\^=7}

\def\subon{\catcode`\_=8}

\def\supsubon{\supon \subon}

\newcommand\actcharon{\catcode`\~=13}

\newcommand\paramon{\catcode`\#=6}

\comment{And now to turn us totally on and off...}

\newcommand\reservedcharson{ \commenton  \mathshifton  \atabon  \supsubon 
	\actcharon  \paramon}

\catcode`@=12
\reservedcharson

\if\papercls \apjcls

\else

\fi

\if\papercls \othercls
\else
\newcommand\inpress{n}
\if\inpress y
\received{\today}
\revised{}
\accepted{}
\if\papercls \apjcls
\slugcomment{}
\fi
\else
\slugcomment{\tablebox{In preparation for {\em ApJ}. DRAFT of {\today}.}}
\fi
\fi

\newcommand\chisq{\ifmmode{\chi\sp{2}}\else\math{\chi\sp{2}}\fi}
\newcommand\redchisq{\ifmmode{ \chi\sp{2}\sb{\rm red}}
	\else\math{\chi\sp{2}\sb{\rm red}}\fi}
\newcommand\Teq{\ifmmode{T\sb{\rm eq}}\else$T$\sb{eq}\fi}
\newcommand\mjup{\ifmmode{M\sb{\rm Jup}}\else$M$\sb{Jup}\fi}
\newcommand\rjup{\ifmmode{R\sb{\rm Jup}}\else$R$\sb{Jup}\fi}
\newcommand\msun{\ifmmode{M\sb{\odot}}\else$M\sb{\odot}$\fi}
\newcommand\rsun{\ifmmode{R\sb{\odot}}\else$R\sb{\odot}$\fi}
\newcommand\mearth{\ifmmode{M\sb{\oplus}}\else$M\sb{\oplus}$\fi}
\newcommand\rearth{\ifmmode{R\sb{\oplus}}\else$R\sb{\oplus}$\fi}

\usepackage[pdftex]{}
\usepackage{epstopdf}
\usepackage{tabularx} 
\usepackage{graphicx}
\usepackage{subfigure}
\usepackage{array,longtable}
\usepackage{float}
\usepackage{amsmath,amsthm,amssymb}

\usepackage{totcount}

\newtotcounter{citnum} 
\def\oldbibitem{} \let\oldbibitem=\bibitem
\def\bibitem{\stepcounter{citnum}\oldbibitem}

\newcolumntype{L}[1]{>{\raggedright\let\newline\\\arraybackslash\hspace{0pt}}m{#1}}
\newcolumntype{C}[1]{>{\centering\let\newline\\\arraybackslash\hspace{0pt}}m{#1}}
\newcolumntype{R}[1]{>{\raggedleft\let\newline\\\arraybackslash\hspace{0pt}}m{#1}}
\newcolumntype{K}[1]{>{\centering\arraybackslash}p{#1}}

\newcommand{\rmn}[1]{\rm{#1}}

\newcommand{\Msun}{\rm{M}_{\odot}}
\newcommand{\Mbh}{\rm{M}_{\rmn{bh}}}
\newcommand{\mstar}{\rm{M}_{\star}}
\newcommand{\mbh}[1]{\rm{M}_{\rmn{bh}_{#1}}}

\newcommand{\Rsun}{\rm{R}_{\odot}}
\newcommand{\rtau}{\rm{R}_{\tau}}
\newcommand{\rstar}{\rm{R}_{\star}}

\newcommand{\rp}{\rmn{R}_{\rm{p}}}
\newcommand{\rl}{\rmn{R}_{\rm{L}}}
\newcommand{\rcmb}{\rmn{R}_{\rmn{c, mb}}}
\newcommand{\rcp}{\rmn{R}_{\rmn{c, p}}}

\newcommand{\amb}{a_{\rm{mb}}}

\newcommand{\aninety}{a_{90}}

\newcommand{\qcrit}{q_{\rmn{crit}}}

\newcommand{\ceff}{\chi_{\rm{eff}}}

\newcommand{\mdot}{\dot{\rmn{M}}}

\newcommand{\jbin}{{\bf J}_{\rm bin}}
\newcommand{\jdisk}{{\bf J}_{\rm disk}}

\shorttitle{Spin Evolution of LIGO Sources}
\shortauthors{Lopez Jr. et al.}

\begin{document}

\title{TIDAL DISRUPTIONS OF STARS BY BINARY BLACK HOLES: MODIFYING THE SPIN MAGNITUDES AND DIRECTIONS OF LIGO SOURCES IN DENSE STELLAR ENVIRONMENTS}

\author{Martin Lopez Jr.\altaffilmark{1}, Aldo Batta\altaffilmark{1,2},
Enrico Ramirez-Ruiz\altaffilmark{1,2},
Irvin Martinez\altaffilmark{2},
Johan Samsing\altaffilmark{3}
}

\affil{
	\sp{1} Department of Astronomy and Astrophysics, University of California, Santa Cruz, CA 95064, USA \linebreak	
	\sp{2} Niels Bohr Institute, University of Copenhagen, Blegdamsvej 17, 2100 Copenhagen, Denmark \linebreak
	\sp{3} Department of Physics \& Astronomy, Princeton University, Princeton, NJ, 08544}

\begin{abstract}
Binary black holes (BBHs) appear to be widespread  and are able to merge through the emission of gravitational waves, as recently illustrated by LIGO.
The spin of the BBHs is one of the parameters that LIGO can infer from the gravitational wave signal and can be used to
constrain their production site. If BBHs are assembled in stellar clusters they are  likely to interact with stars,
which could occasionally lead to a tidal disruption event (TDE). When a BBH tidally disrupts a star it can accrete a significant fraction of the debris, effectively altering the spins of the BHs. Therefore, although dynamically formed BBHs are expected to have random spin orientations, tidal stellar interactions can significantly alter their birth spins both in direction and magnitude. Here we investigate how TDEs by BBHs can affect the properties of  the  BH members as well as exploring the characteristics of the resulting electromagnetic signatures.  We conduct hydrodynamic simulations with a Lagrangian Smoothed Particle Hydrodynamics code of a wide range of representative tidal interactions.  We find that both spin magnitude and orientation can be altered and temporarily aligned or anti-aligned through accretion of stellar debris, with a significant dependence on the mass ratio of the disrupted star and the BBH members. These tidal interactions feed material to the BBH at very high accretion rates, with the potential to launch a relativistic jet. The corresponding beamed emission is a beacon to an otherwise quiescent BBH.

\end{abstract}

\keywords{black holes, tidal disruptions, close binaries, dense stellar systems, LIGO}


\section{INTRODUCTION}
\label{intro}

A watershed event occurred on September 14 2015, when the Laser Interferometer Gravitational-Wave Observatory (LIGO) succeeded in detecting the first  gravitational wave (GW) signal \citep{2016PhRvL.116f1102A}, GW150914, of a binary black hole (BBH) merger. This detection, followed by five others, has unveiled a population of stellar mass BHs  that is significantly heavier than those inhabiting X-ray binaries \citep{2011ApJ...741..103F}.

A large number of progenitor systems have been suggested, all designed to manufacture  BHs in the observed mass range. The two most widely discussed scenarios encompass dynamical assembly in dense star clusters \citep{1993Natur.364..423S,2000ApJ...528L..17P,2010MNRAS.407.1946D,2011MNRAS.416..133D,2014MNRAS.441.3703Z,2014ApJ...784...71S,2015PhRvL.115e1101R,2016ApJ...832L...2R,2016ApJ...824L...8R,2017ApJ...840L..14S,2018ApJ...855..124S}
and isolated massive stellar field binaries \citep{1976IAUS...73...75P,1993PASP..105.1373I,2001ASPC..229..239P,2003MNRAS.342.1169V,2007PhR...442...75K,2010NewAR..54...65T,2012ApJ...759...52D,2013ApJ...779...72D,2013A&ARv..21...59I,2014LRR....17....3P,2016Natur.534..512B, 2018ApJ...862L...3S}, including chemically homogeneous stars \citep{2009A&A...497..243D,2016A&A...588A..50M,2016MNRAS.458.2634M,2016MNRAS.460.3545D}.

Other proposed scenarios include active galactic nuclei (AGN) discs \citep{2017ApJ...835..165B,  2017MNRAS.464..946S, 2017arXiv170207818M},
galactic nuclei \citep{2009MNRAS.395.2127O, 2015MNRAS.448..754H,
	2016ApJ...828...77V, 2016ApJ...831..187A, 2016MNRAS.460.3494S, 2017arXiv170609896H},
single-single GW captures of primordial BHs \citep{2016PhRvL.116t1301B, 2016PhRvD..94h4013C,
	2016PhRvL.117f1101S, 2016PhRvD..94h3504C}, and very massive stellar mergers \citep{Loeb:2016, Woosley:2016, Janiuk+2017, DOrazioLoeb:2017}.
Generally, these theoretical predicted channels can be broadly tuned to  be consistent with the properties and rates of the BBH sources observed by LIGO so far,
and the challenge remains  to find reliable observational tests. 

Recent work suggests that the key parameters that might help discriminating between formation channels include
the BH mass \citep[e.g.][]{2017ApJ...846...82Z}, orbital eccentricity
in LIGO \citep{2009MNRAS.395.2127O, 2012PhRvD..85l3005K, 2014ApJ...784...71S, 2016ApJ...824L..12O, 2017ApJ...840L..14S, 2018MNRAS.476.1548S, 2018ApJ...853..140S, 2019MNRAS.482...30S, 2018PhRvD..97j3014S, 2018ApJ...855..124S, 2018arXiv181000901Z, 2018arXiv181104926R, 2018ApJ...860....5G} and
LISA  \citep[\textit{e.g.},][]{2018MNRAS.tmp.2223S}, and especially the dimensionless spin parameter $\chi_{\rm{eff}}$ \citep{2018ApJ...854L...9F,2017Natur.548..426F,2016ApJ...832L...2R,2018ApJ...862L...3S}.
$\chi_{\rm{eff}}$ is the total mass weighted BH spin components in the direction of the orbital angular
momentum,
\begin{align} 
\ceff = \frac{\mbh{1} \mathbf{a}_{{\rm bh}_1}+ \mbh{2} \mathbf{a}_{{\rm bh}_2}}{\mbh{1} + \mbh{2}} \cdot \mathbf{\hat{L}}.
\end{align}
Here $\mathbf{a}_{{\rm bh}_1}$ and $\mathbf{a}_{{\rm bh}_2}$ are the dimensionless spins of the BHs and $\mathbf{\hat{L}}$ is the direction of the orbital angular momentum. The spin measurements of BBHs arising from the isolated massive stellar field binary scenario roughly predicts  alignment of the BH spins and the orbital angular momentum  \citep{2000ApJ...541..319K}, while dynamically assembled BHs  are expected to have uncorrelated spins as they are formed and harden through a series of chaotic exchange interactions \citep{2016ApJ...832L...2R}.

Here we will analyze the dynamical scenario and investigate whether the determination of $\ceff$  allows for constraints to be placed on the spin history of the BBH
system between assembly and merger. Such a BBH becomes detectable only through
interactions with its gaseous  environment. Gas that is lost from nearby stars, or even stars plunging into such binaries, can produce detectable signatures. Through the use of Smoothed Particle Hydrodynamic (SPH) simulations, we show how stellar material which is accreted following a tidal disruption event (TDE) can alter the birth spin magnitudes and orientation of the individual BHs, possibly aligning or misaligning them temporarily. Furthermore, the supply of material to the BBH is above the Eddington limit and could launch a relativistically-beamed jet. The emerging class of high energy transient bursts all have peak luminosities and
durations reminiscent of ultra-long $\gamma$-ray bursts. Tidal disruptions of stars by BBHs thus uniquely probe the currently-debated existence of LIGO signals emanating from dense star clusters. 

The structure of the paper is as follows. Section \ref{sec:TDEs} discusses the dynamics of LIGO BBH (LBBH) TDEs in dense star clusters. Section \ref{sec:hydro} overviews the hydrodynamic formalism and  presents the results as well as their significance for the spin magnitude and alignment of the individual BHs. Section \ref{sec:disc} explores the implications of our results and possible sources for upcoming high energy transient surveys.


\section{Tidal Disruption Events by LIGO BBHs}
\label{sec:TDEs}

\begin{figure*}[t!]
	\includegraphics[width=\textwidth]{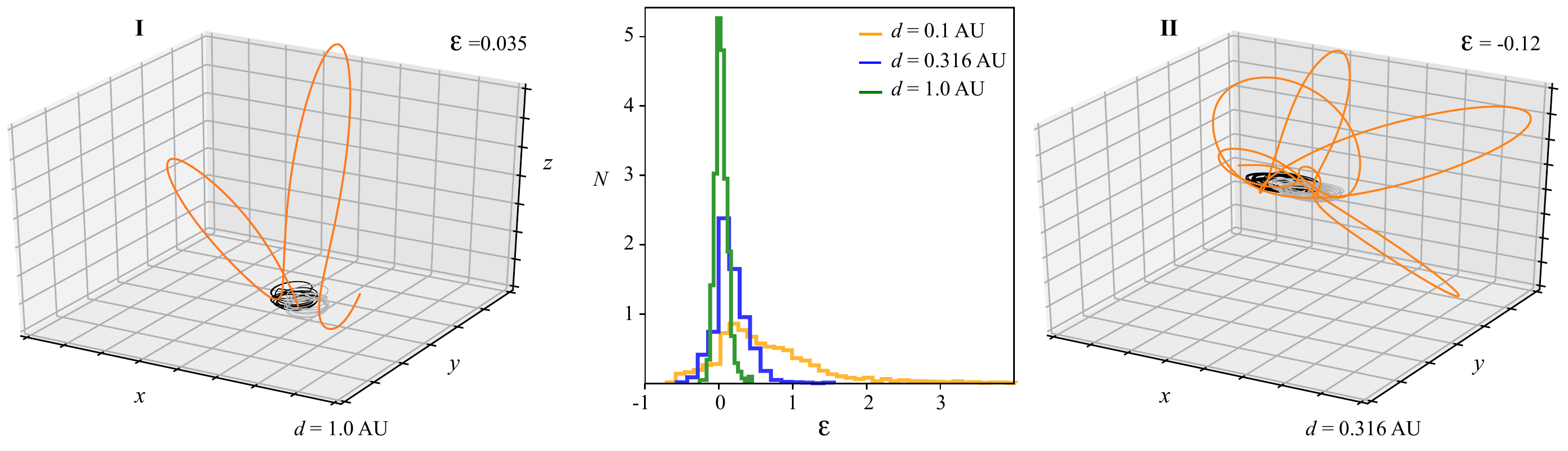}
	\caption{The CM energy distributions with respect to the disrupting BH and specific trajectories are shown for a sun-like star ($M_\ast=1M_\sun$, $R_\ast=R_\sun$) interacting with a $15 \Msun$ equal mass BH binary with $e = 0.5$. The properties of the binary have been selected to reflect those derived by \citet{2016PhRvD..93h4029R} for dynamically assembled LBBHs. Here we study the outcomes of TDE interactions and their associated CM energy distributions by performing large set of numerical scattering experiments using the $N$-body code developed by \citet{2014ApJ...784...71S}. The  panels  shows the energy distribution for different binary separations ($d=1.0 \ \rmn{AU} = 87.3 \rtau$, $d=0.316 \  \rmn{AU} =27.6\rtau$, $d=0.1 \ \rmn{AU} =8.73 \rtau$) and the trajectories of unbound (I) and bound (II)  stellar orbits ({\it orange} trajectories) . Here $\varepsilon$ is the CM energy of the star with respect to the disrupting BH at $\rtau$ in units of the binding energy of the star.}
	\label{fig:stats}
\end{figure*}

\subsection{Single BH Dynamics}
\label{subsec:dynamics}
Canonical TDEs occur when a star with mass $M_\ast$ and radius $R_{\ast}$ gets disrupted when approaching a  supermassive black hole  (SMBH) with mass $M_{\rmn{bh}}$  at a pericenter  distance $R_{\rm p}=R_{\tau} = q^{-1/3}R_{\ast}$, where $q=M_\ast / M_{\rmn{bh}}$ \citep{1988Natur.333..523R,1989IAUS..136..543P,1989ApJ...346L..13E}. After the disruption, about half of the star becomes unbound and ejected, while the other half
becomes bound to the SMBH on elliptical orbits. 3D hydrodynamical simulations have quantified the rate at which material falls back onto the SMBH \citep{2013ApJ...767...25G}.
A good  fit to observed light curves of TDEs is obtained if one assumes
that the accretion luminosity  directly follows the fallback rate  in the simulation \citep{2018arXiv180108221M}.
However, it is not clear why this should be the case. Bound debris returns to the SMBH with a large range of eccentricities and orbital periods \citep{2009ApJ...697L..77R}
and it may take many Keplerian orbits for fallback material to circularize and accrete \citep{2015ApJ...809..166G}. Some mechanism is therefore required  to quickly dissipate the kinetic energy of the fallback material and circularize it into an accretion disk.

In standard TDE discourse \citep{1988Natur.333..523R}, the disrupting SMBHs have masses $M_{\rmn{bh}}\gtrsim 10^6 M_{\odot}$ yielding $q \ll 1$, which  allows the semi-major axis of the most bound material to be approximated as:
\begin{align}
\amb &=  \left(\frac{\Mbh}{\mstar}\right)^{1/3} \rtau \nonumber\\
&=  q^{-1/3} \ \rtau.
\label{eq:amb_old}
\end{align}
However for disrupting BHs within a LBBH, the mass ratio is near unity, making the extent of the star comparable to the tidal radius. In this case, the specific orbital energy of stellar material varies significantly across the star:
\begin{align}
E(r) &= -G {\Mbh} \left[\sum_{n = 1}^{\infty} \left(\frac{q^{1/3}}{\rstar}\right)^{n+1}  r^n\right],
\end{align}
where $r$ is the distance from the star's center of mass (CM). For material that is bound to the BH, this expression translates into a  range of semi-major axes  given by:
\begin{align}
a(r) &= -\frac{G\Mbh}{2E(r)} \nonumber \\
&= \left[2 \sum\limits_{n=1}^{\infty}\left(\frac{q^{1/3}}{\rstar}\right)^{n+1}  r^n\right]^{-1},
\label{eq:semi_material}
\end{align}
which for canonical TDEs $\left(q \ll 1\right)$ can be safely approximated to first order. As $q$ approaches unity this approximation is no longer valid and the semi-major axis of the most bound material approaches the tidal radius and becomes equal to it at a critical mass ratio $\qcrit =0.037$. The assumption that the circularization radius of the most bound material is about twice the tidal radius \citep{1990ApJ...351...38C,1998AIPC..431..141U,2009ApJ...698.1367G,2011MNRAS.410..359L,2011MNRAS.415..168S,2014ApJ...783...23G} also breaks down in the LBBH regime. The circularization radius  of the most bound material $\rcmb$ in this case  is given by
\begin{align}
{\rcmb} = 2 {\rtau} \left[1 - q^{1/3}\right]^2,
\label{eq:rcmb}
\end{align}
while the spread in circularization radii can be written as 
\begin{align}
\frac{\Delta {\rmn{R}_{\rmn{c}}}}{\rcp} = q^{1/3} \left[2 - q^{1/3}\right],
\end{align}
where the circularization radius of the pericenter is $\rcp=2\rtau$.
In order for this material to circularize and form a disk, energy must be dissipated efficiently after disruption.  
Material falling to pericenter can be heated by hydrodynamical shocks and \cite{2014ApJ...783...23G} show that the fractional energy dissipation per orbit, $\nu_{\rmn{H}}$, can be written as 
\begin{align}
\nu_{\rmn{H}} = \beta q^{2/3}
\label{eq:diss}
\end{align}
where $\beta = \rp/\rtau$. For disruptions in the LBBH regime, the energy dissipation via shocks at pericenter can be sizable  and   lead to efficient circularization.
This is in contrast to the standard case with $q \ll 1$,  for which 
hydrodynamical  shocks  at pericenter are likely to be insufficient and rapid circularization might only be achieved via general
relativistic effects  \citep{2015ApJ...804...85S,2015ApJ...809..166G,2016MNRAS.455.2253B,2016MNRAS.461.3760H}. We note here that not all the binaries we refer to as  LBBH will  necessarily merge.

\subsection{Binary BH Dynamics}
\label{subsec:BBHdyn}
For BBH TDEs, the star does not necessarily follow a parabolic orbit  and the orbital deviations before disruption depend strongly on the separation $d$ and eccentricity $e$ of the binary. 
The CM energy distributions of a sun-like star with respect to the disrupting BH part of a $15  \Msun$ equal mass BBH with $e = 0.5$ are shown in Figure $\ref{fig:stats}$ for three distinct binary separations.  In this case $\rtau=2.47 \rstar =0.01 \ AU$. For ${\rtau}/d \ll 1$ the CM energy is essentially parabolic while a larger fraction of unbound CM orbits are observed for tighter binaries. This is partly due to the individual BHs evolving faster around their binary CM as BBHs get tighter, which then maps to a higher
relative velocity at the time of disruption and thereby a higher relative energy. Note here that stellar elements unbound with respect
to the disrupting BH can still be bound to the CM of the BBH, which then can lead to later accretion.

After disruption, the fate of the debris also depends sensitively on the ratio ${\rtau}/d$. If ${\rtau}/d > 1$, the disruption will take place outside of the binary and the infalling material will form a circumbinary disk around the system. In what follows, we refer to this scenario as the \textit{circumbinary scenario} (CS).   When ${\rtau}/d \lesssim 1$, the star will be disrupted by one of the  binary members but the accretion history  of the debris onto the system  is determined by $d$. This is due to the debris orbiting around the disrupting BH with a wide range of semi-major axes such that there is  always some material  that is able to reach the sphere of influence of the companion BH. In order to determine whether  or not the non-disrupting BH  can  accrete  significant  amounts of stellar debris we make  use of two important  characteristic  scales.  One is the semi-major axis $\aninety$ of the disrupted material whose orbit contains 90\% of the stellar debris. In other words, $\aninety$ is the semi-major of material whose radius, measured from the most bound material of the star inwards, contains $90\%$ of the stellar mass. Therefore we classify a strong interaction as being one where the non-disrupting BH interacts with $10\%$ of stellar debris.
The other scale is the Roche lobe radius $\rl$, which determines the gravitational sphere of influence of the disrupting BH. $\rl$ can be  written \citep{1983ApJ...268..368E} as 
\begin{align}
{{\rl} \over d_{\rmn{min}}} = \frac{0.49 {q_{\rm b}}^{2/3}}{0.6 q_{\rm b}^{2/3} + \ln{\left(1 + q_{\rm b} ^{1/3}\right)}},
\label{eq:roche}
\end{align}
where $q_{\rm b}$ is the mass ratio of the BBH and $d_{\rmn{min}}$ is the minimum separation of the binary. When $\aninety/\rl < 1$, a small fraction of the debris is able to  interact with the non-disrupting BH but most of the stellar debris will be accreted by the disrupting BH. In this case, the tidal interaction  will resemble that caused  by a single BH and we  refer to this as  the \textit{single scenario} (SS). On the other hand, disrupted material with $\aninety/\rl \gtrsim 1$ will be influenced by the companion and a sizable  fraction of debris can be accreted by the non-disrupting BH. A case we refer to as the \textit{overflow scenario} (OS). 

In order to calculate the spin change due to accretion of disrupted material we use  \citep{1970Natur.226...64B}
\begin{equation}
\resizebox{.9 \linewidth}{!} 
{
	$ S(\rmn{M}_{\rm bh,f}) = \left(\frac{2}{3}\right)^{1/2} \frac{\Mbh}{\rmn{M}_{\rm bh,f}} \left\{4-\left[18\left(\frac{\Mbh}{\rmn{M}_{\rm bh,f}}\right)^{2} - 2 \right]^{1/2}\right\},$ 
}
\label{eq:spin}
\end{equation}

which assumes an initially low or non-spinning BH. Here $\rmn{M}_{\rm bh,f}=\rmn{M}_{\rm bh}+f\mstar$ is the final mass of the BH after accreting a fraction $f$ of the disrupted star. For a TDE of a  star in a  parabolic orbit ($f=0.5$), the maximum mass that the BH can accrete is $0.5 \mstar$ such that the maximum spin up, $S_{\rmn{max}}$ is given by
\begin{equation}
\resizebox{.85 \linewidth}{!} 
{
	$S_{\rmn{max}}(q) = \left(\frac{8}{3}\right)^{1/2} \left(\frac{1}{2 + q}\right) \left\{4-\left[72\left(\frac{1}{2 + q}\right)^{2} - 2 \right]^{1/2}\right\},$
	\label{eq:maxspin}
}
\end{equation}
The values of $S_{\rmn{max}}$ for a few characteristic $q$'s are $S_{\rmn{max}} \left(q = 1 \times 10^{-6} \right) = 1.84 \times 10^{-6}$, $S_{\rmn{max}}\left(q=0.01\right) = 0.02$, and $S_{\rmn{max}}\left(q = 0.5\right) = 0.60$. This clearly illustrates that for LBBHs, the digestion of stars during the lifetime of the binary could lead to noticeable spin  changes. 


\section{Hydrodynamics}
\label{sec:hydro}

\begin{figure*}[b]
	
	\begin{center}
		\begin{minipage}[c]{0.3\textwidth}
			\includegraphics[width=\textwidth]{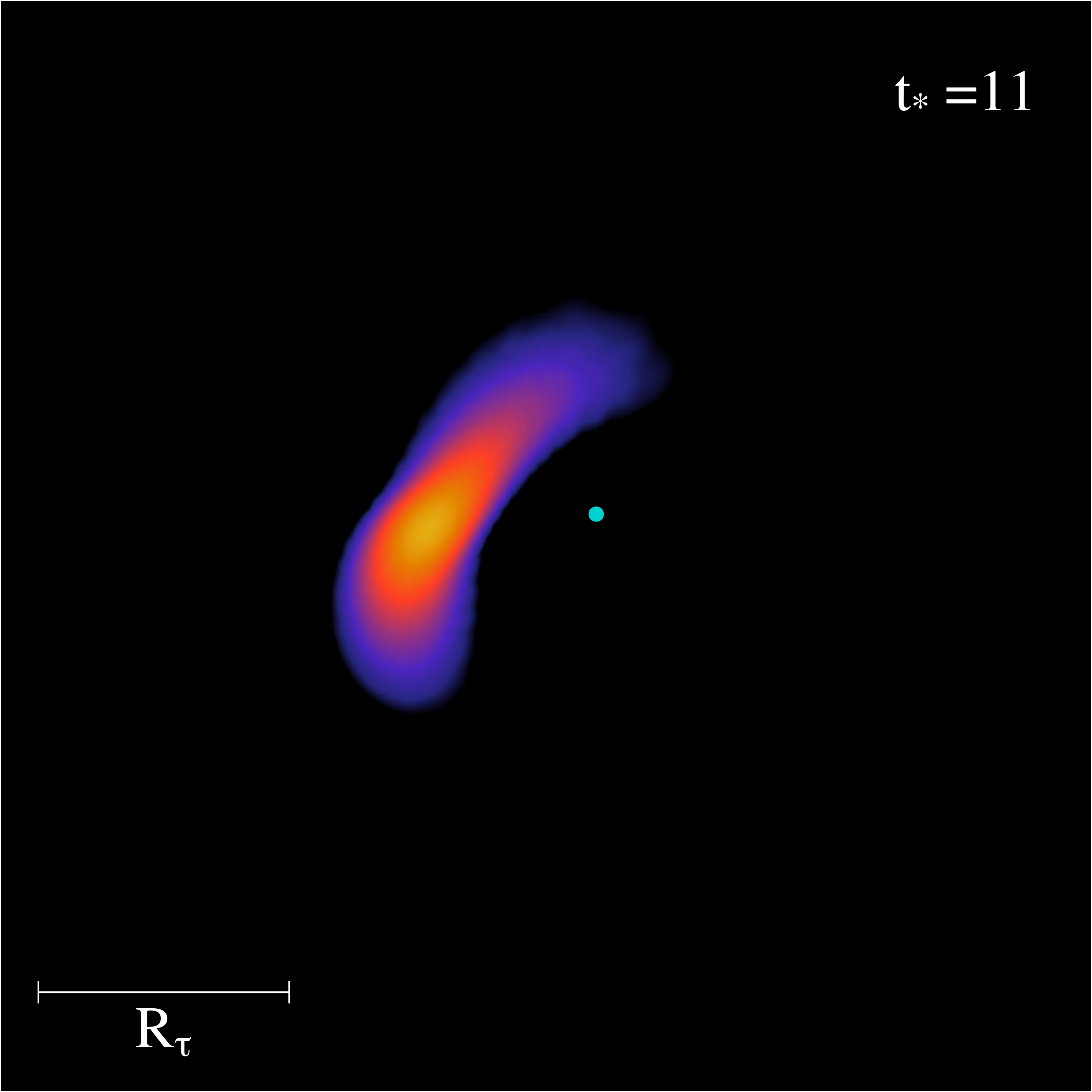}
		\end{minipage}
		\hspace{-0.3cm}
		\begin{minipage}[c]{0.3\textwidth}
			\includegraphics[width=\textwidth]{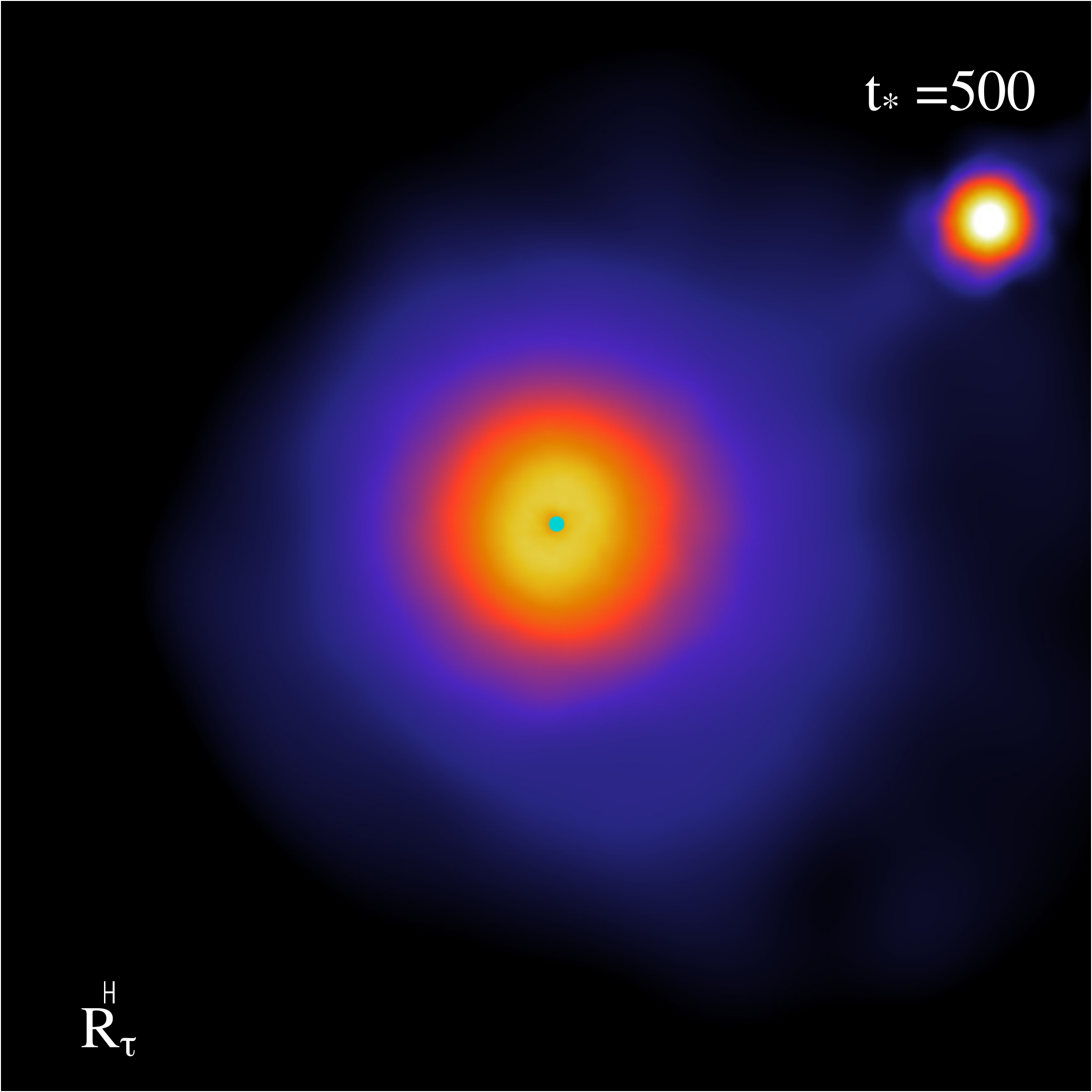}
		\end{minipage}
		\hspace{-0.3cm}
		\begin{minipage}[c]{0.3\textwidth}
			\includegraphics[width=\textwidth]{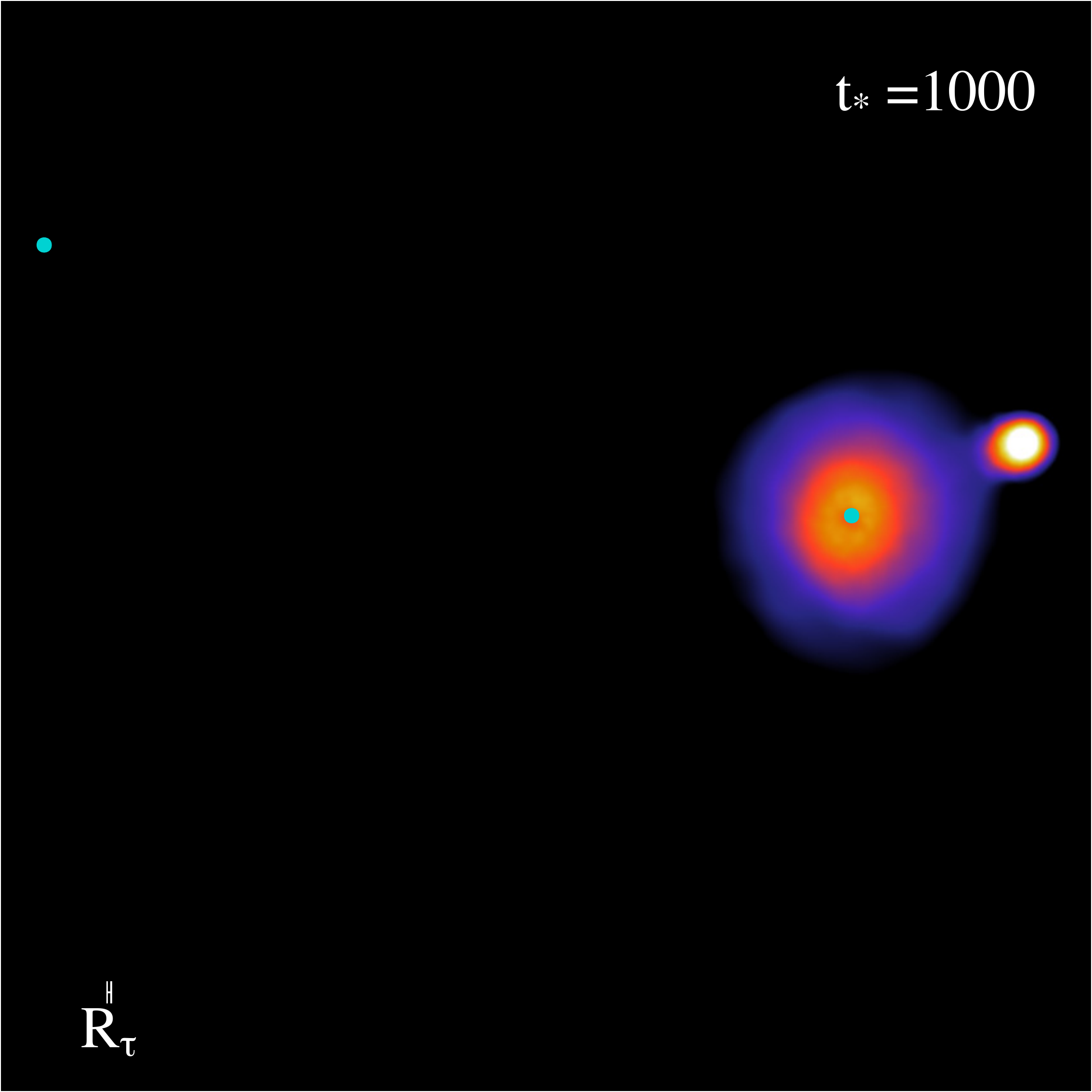}
		\end{minipage}
	\end{center}
	
	\vspace{-0.5cm}
	
	\begin{center}
		\begin{minipage}[c]{0.3\textwidth}
			\includegraphics[width=\textwidth]{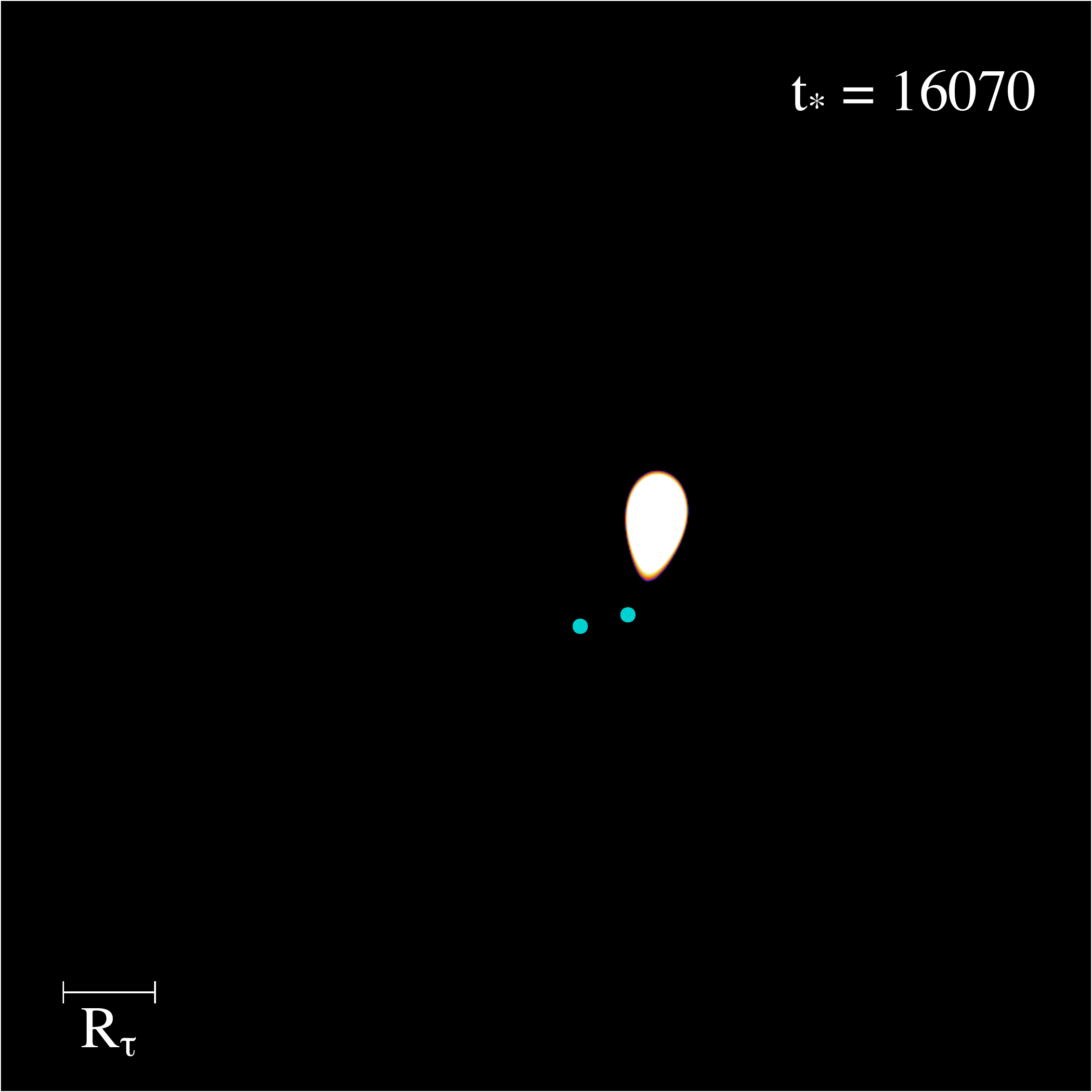}
		\end{minipage}
		\hspace{-0.3cm}
		\begin{minipage}[c]{0.3\textwidth}
			\includegraphics[width=\textwidth]{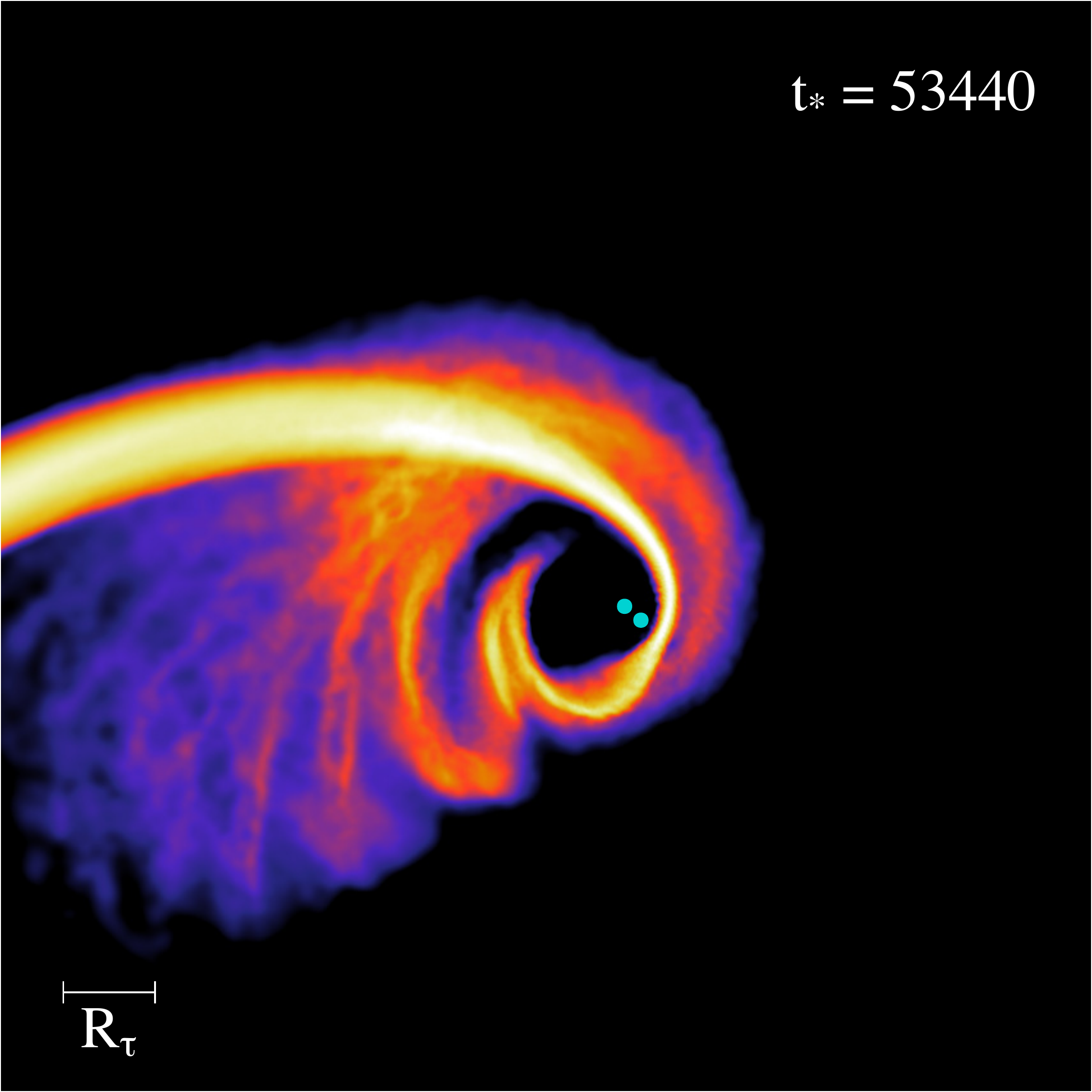}
		\end{minipage}
		\hspace{-0.3cm}
		\begin{minipage}[c]{0.3\textwidth}
			\includegraphics[width=\textwidth]{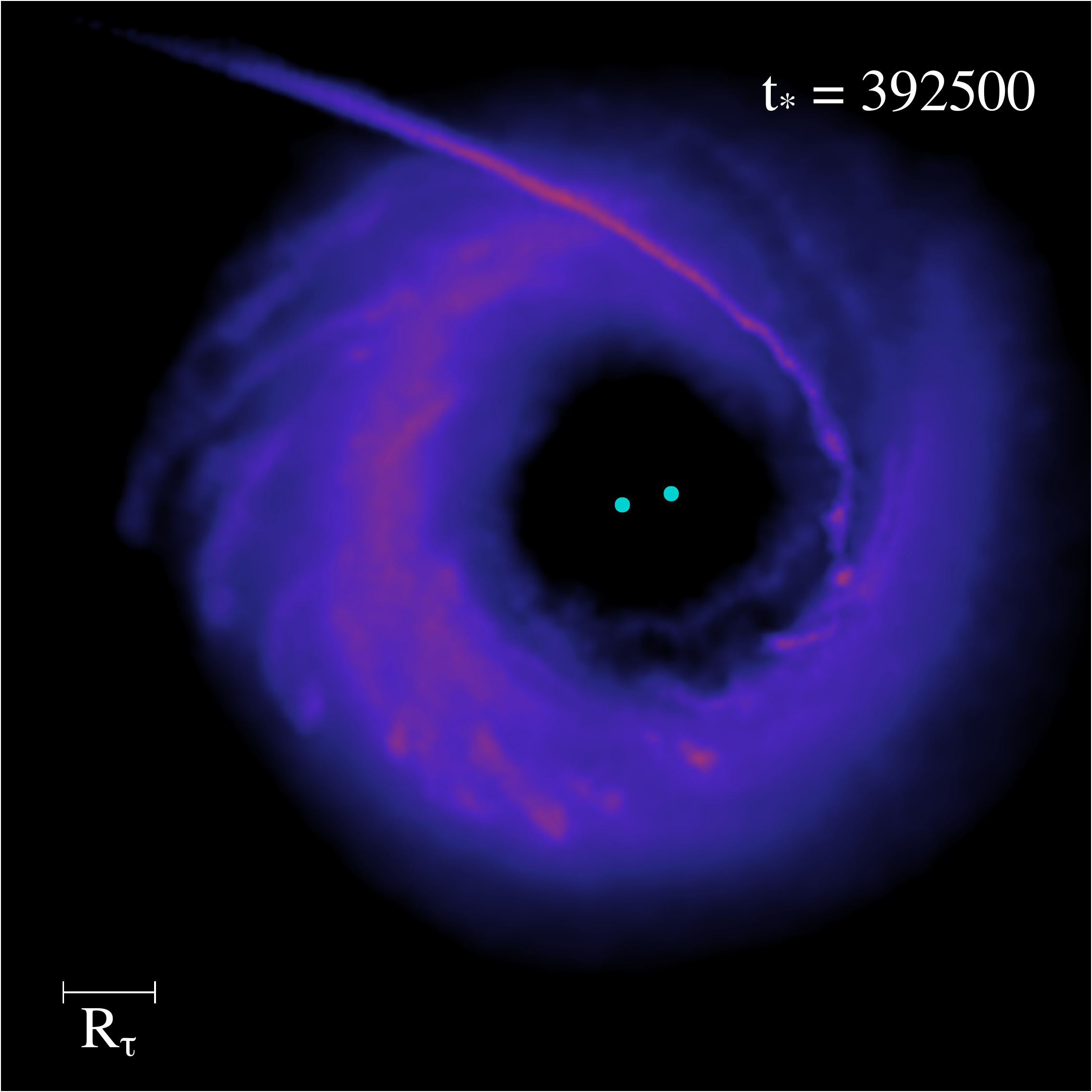}
		\end{minipage}
	\end{center}
	
	\vspace{-0.5cm}
	
	\begin{center}
		\begin{minipage}[c]{0.3\textwidth}
			\includegraphics[width=\textwidth]{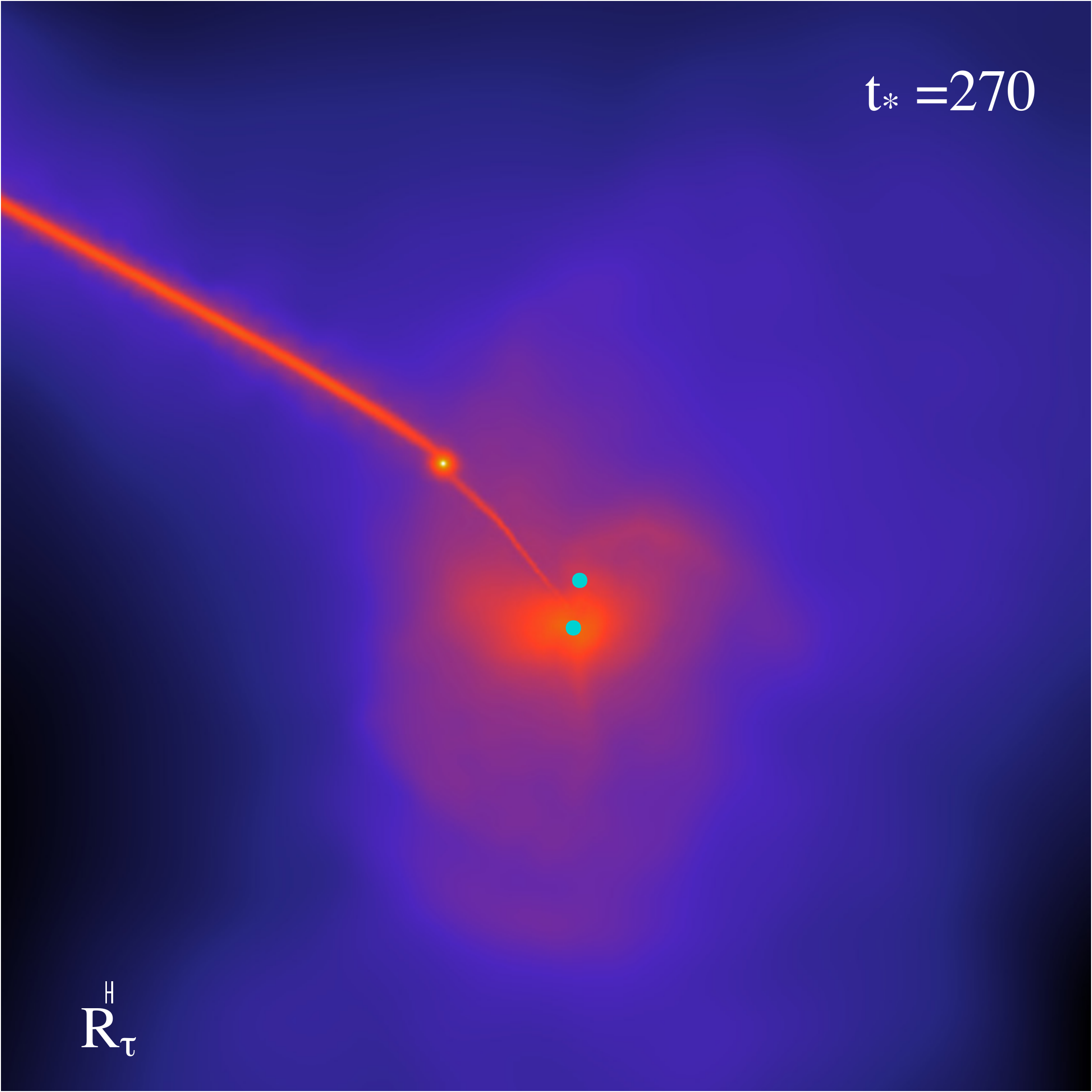}
		\end{minipage}
		\hspace{-0.3cm}
		\begin{minipage}[c]{0.3\textwidth}
			\includegraphics[width=\textwidth]{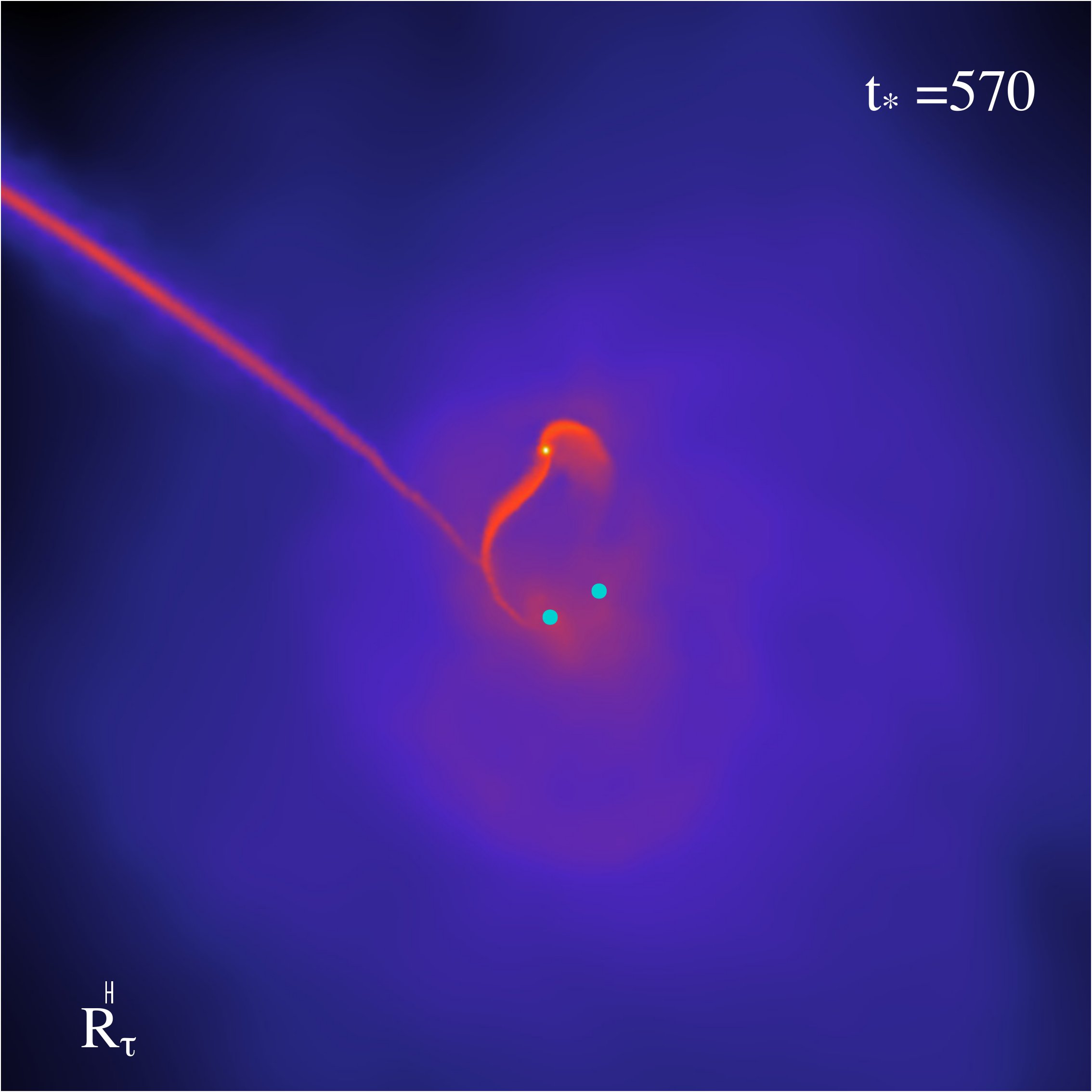}
		\end{minipage}
		\hspace{-0.3cm}
		\begin{minipage}[c]{0.3\textwidth}
			\includegraphics[width=\textwidth]{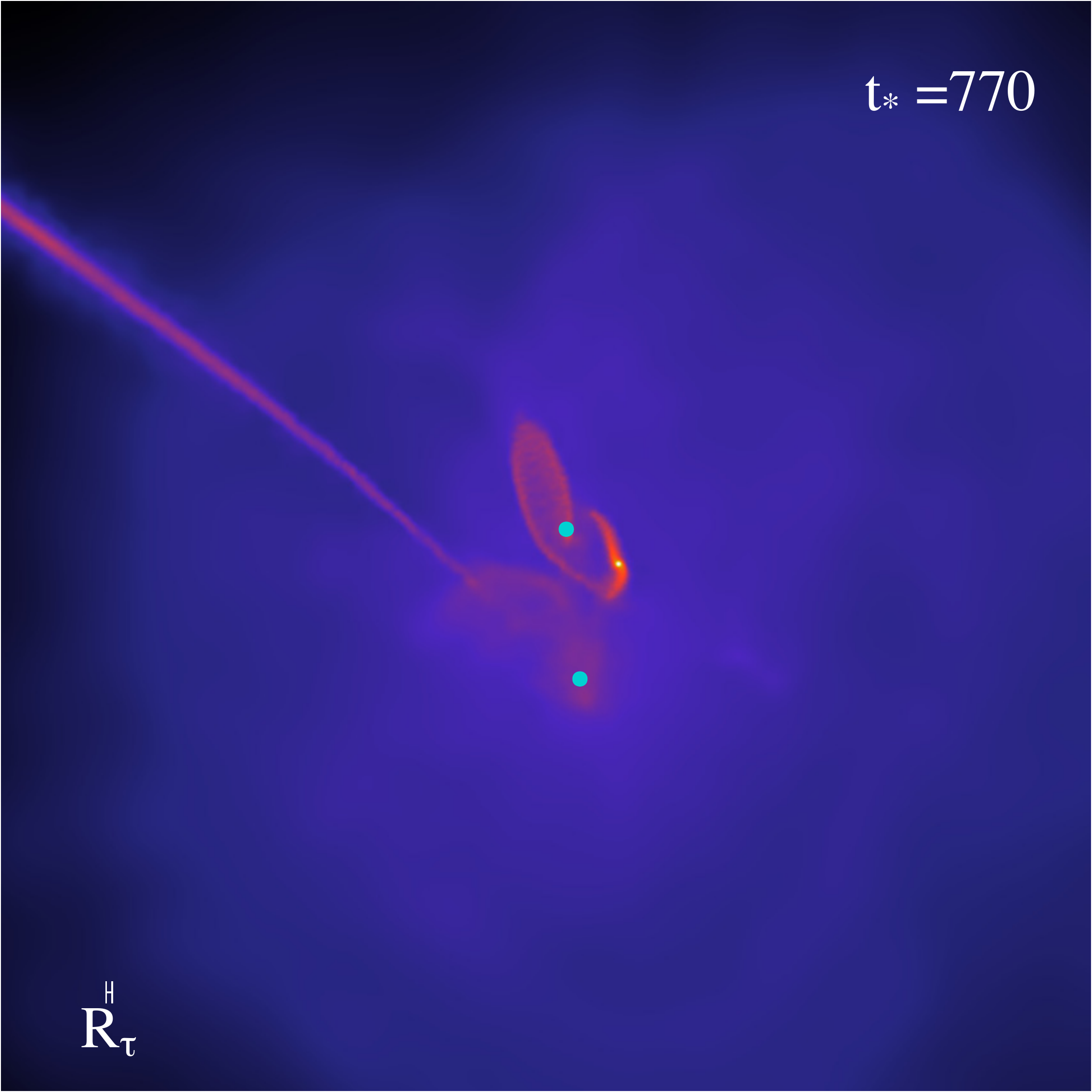}
		\end{minipage}
	\end{center}
	
	\caption{Simulations of the tidal interaction of stars with a LBBH. Here $t_{*}$ and $R_{\tau}$ are the simulation times in dynamical time units and the corresponding tidal radii. All panels are in the orbital plane of the LBBH. {\it Top panels:} Simulation of the SS case at three different times, from disruption to the subsequent accretion onto the disrupting BH. {\it Middle panels:} Simulation of the CS case at three different times, from  the initial disruption occurring outside the LBBH to the assembly of the circumbinary disk. {\it Bottom panels:} Simulation of the OS case, from  initial partial disruption of the star, followed by a second and third disruption of the remaining stellar core. During this interaction, a  total of four disruptions occur. The simulation parameters, listed as [SS, CS, OS], are: $N= [10^5,10^6,10^5]$, $\rstar = [1, 43, 1]\Rsun$, $\Gamma = [4/3, 5/3, 4/3]$, $d = [429.88, 42.99, 42.99]\Rsun$, $v_{\infty} = [30, 10, 20] \rm{km/s}$. In all cases $\mstar = 1\Msun$, $\mbh{1} = \mbh{2} = 15\Msun$, and $e = 0.5$.}
	\label{fig:unique}
\end{figure*}

\subsection{Set-Up}
Our hydrodynamical simulations of LBBH TDEs use a modified version of the SPH code Stellar GADGET-3 \citep{2005MNRAS.364.1105S,2012MNRAS.424.2222P}. GADGET-3 allows one to accurately follow the accretion of material into sink particles and the compressibility of the gas is described with a gamma-law equation of state $P\propto \rho^\gamma$. By solving the Lane-Emden equation and using the same method as in \citet{2017ApJ...846L..15B}, we created three-dimensional  spherically symmetric distributions of SPH particles by mapping polytropic stars in hydrodynamical equilibrium with a structural gamma $\Gamma$ set to either 5/3 or 4/3, representative of low and high-mass stars, respectively. During the simulation, the stars are evolved hydrodynamically according to a $\gamma=5/3$ equation of state, with the difference between $\Gamma$ and $\gamma$ for higher-mass (or convective) stars being a consequence of radiation transfer in the  star's interior. We ran test cases of the tidal disruption of a 1$\Msun$ star by an equal mass $\mbh{1} = \mbh{2} = 15\Msun$ LBBH with varying resolutions between $N=10^{5}$ and $10^{6}$ particles, which showed clear convergence for the accretion rates and mass bound to the system.
\subsection{Initial Conditions}
\label{subsec:IC}

All initial conditions (ICs) assume typical parameters for LBBHs and stars in globular clusters (GCs). We take $e=0.5$ for the LBBH's eccentricity and assume that the individual spins of the BHs ($S_1$ and $S_2$) to be initially zero, which is consistent with the small spins observed for LIGO events so far \citep{2018arXiv181112907T}. 
By means of a three-body code, we obtained the dynamical properties of the LBBH and star prior to a tidal disruption, tracing the trajectories for all three bodies back in the time when the incoming star lies about six tidal radii away from the disrupting BH. These dynamical properties were included in the GADGET-3 IC file. 

\begin{figure*}[t!]
	\begin{center}
		\begin{minipage}[c]{0.24\textwidth}
			\includegraphics[width=\textwidth]{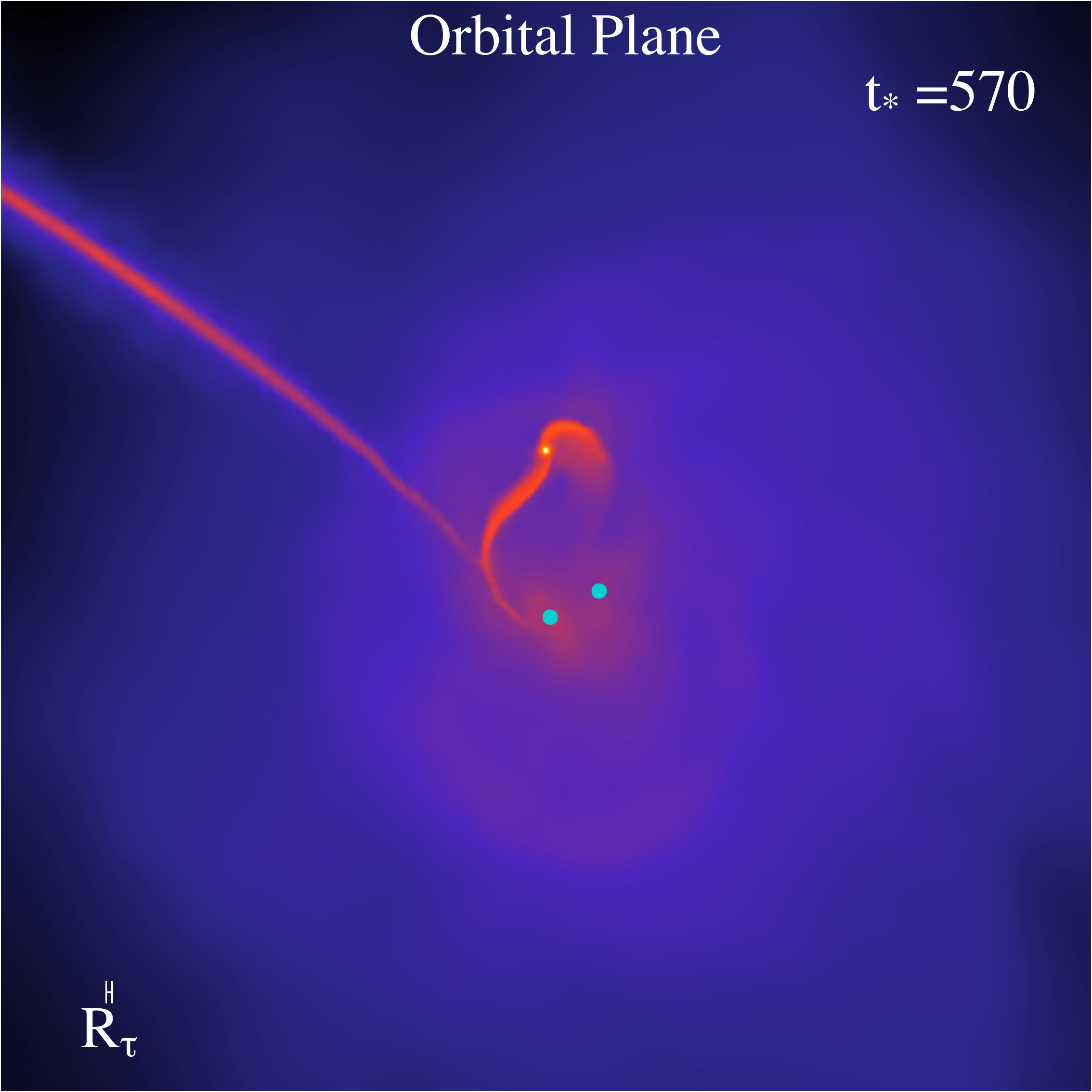}
		\end{minipage}
		\hspace{-0.3cm}	
		\begin{minipage}[c]{0.24\textwidth}
			\includegraphics[width=\textwidth]{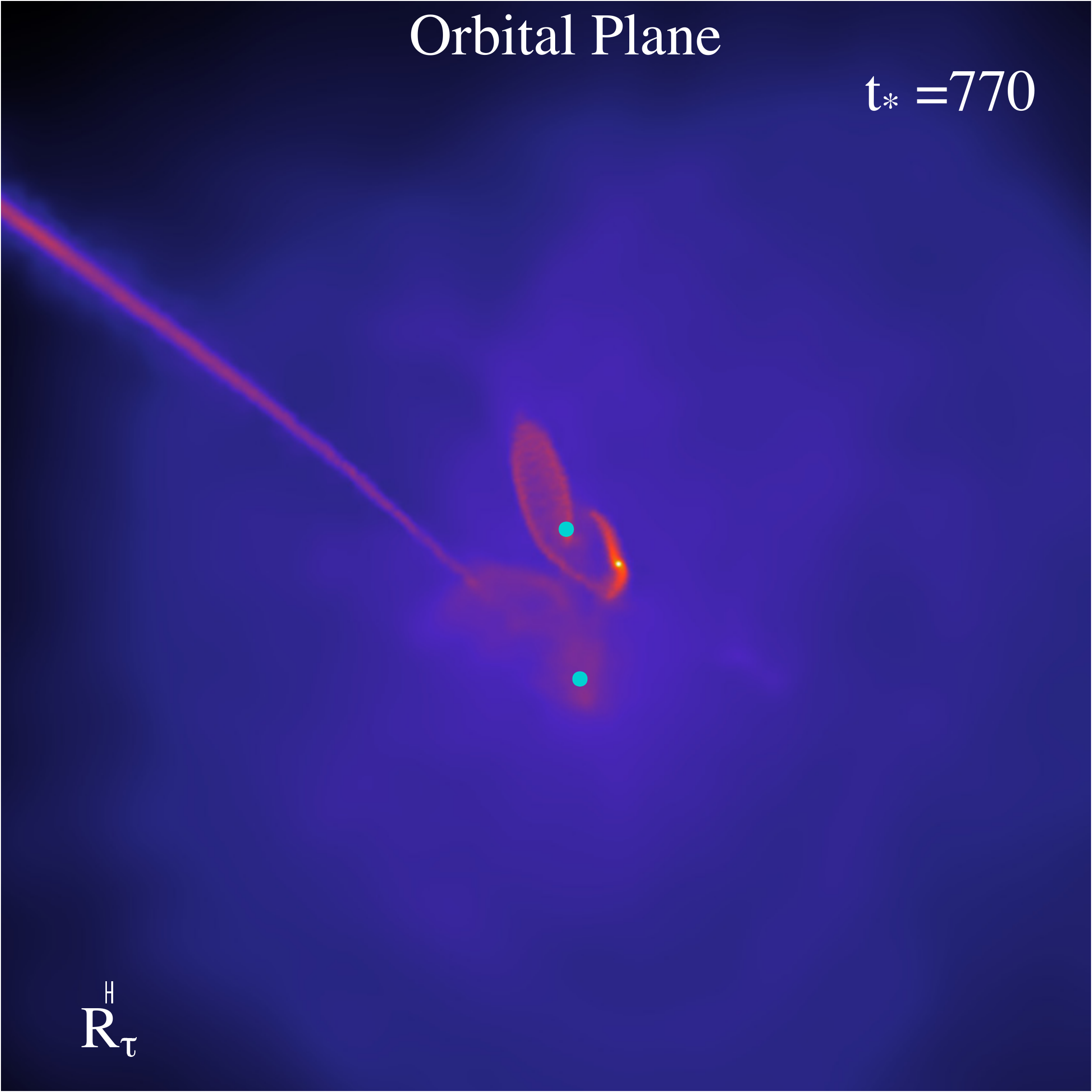}
		\end{minipage}
		\hspace{0.3cm}
		\begin{minipage}[c]{0.24\textwidth}
			\includegraphics[width=\textwidth]{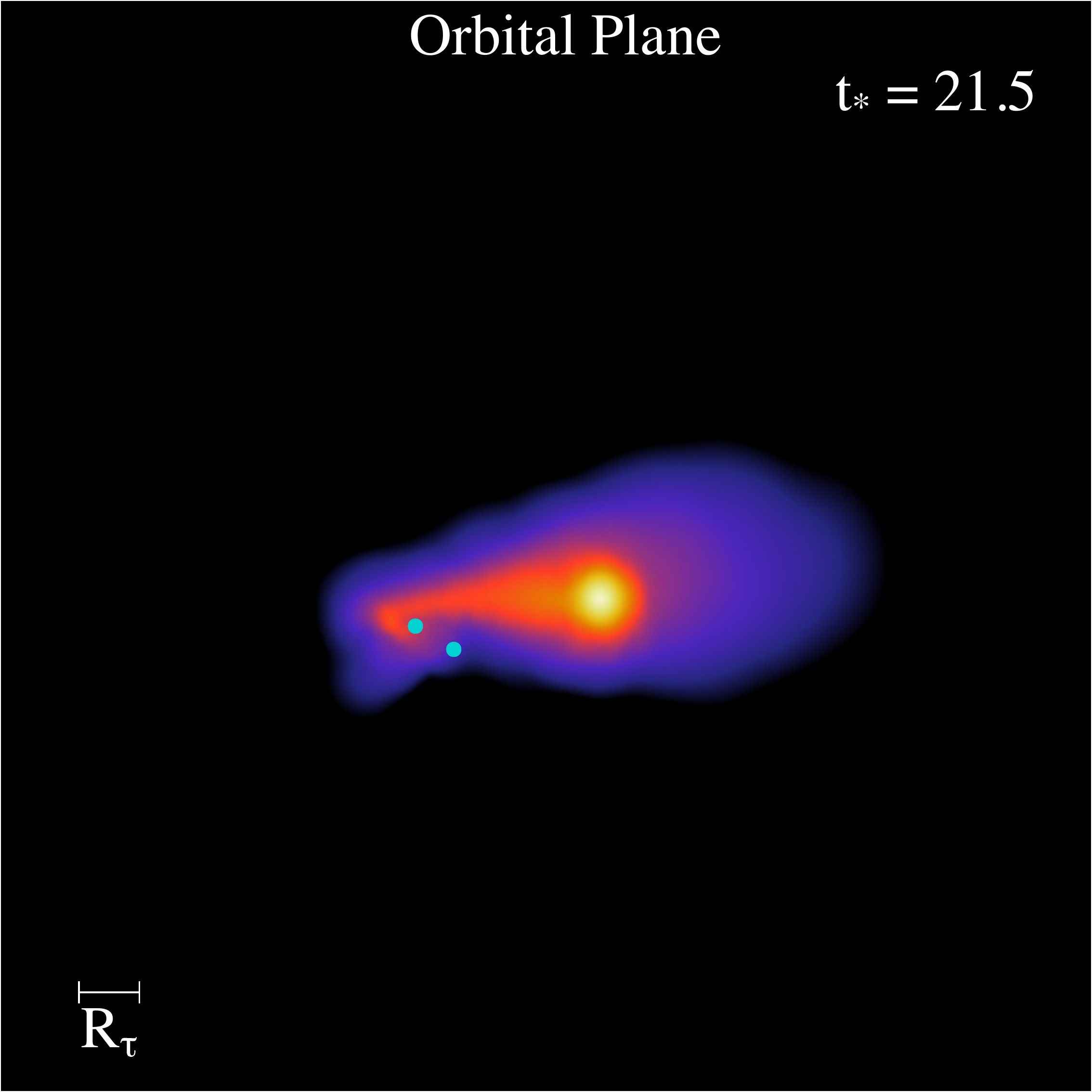}
		\end{minipage}
		\hspace{-0.3cm}
		\begin{minipage}[c]{0.24\textwidth}
			\includegraphics[width=\textwidth]{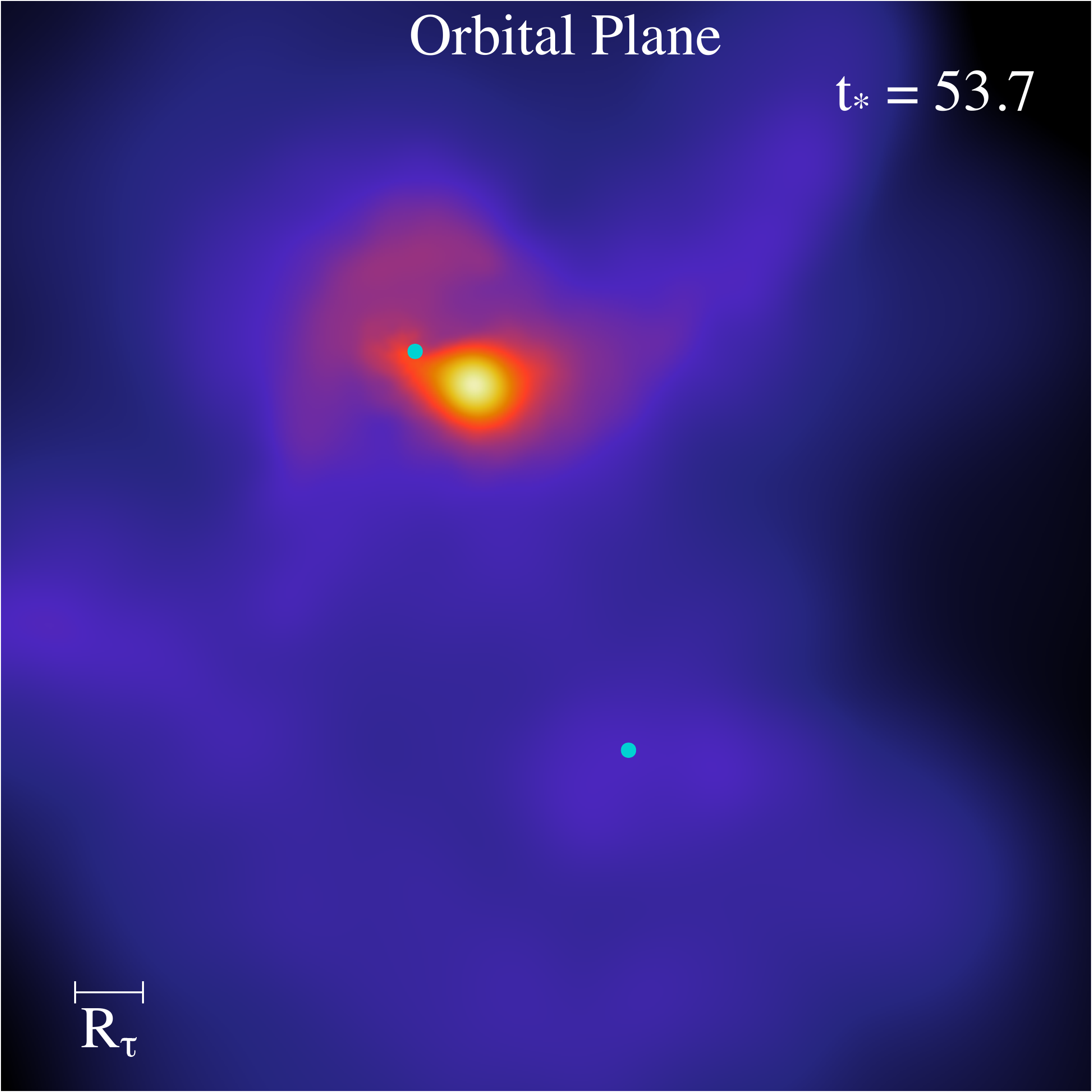}
		\end{minipage}
	\end{center}
	
	\vspace{-0.75cm}
	
	\begin{center}
		\begin{minipage}[c]{0.24\textwidth}
			\includegraphics[width=\textwidth]{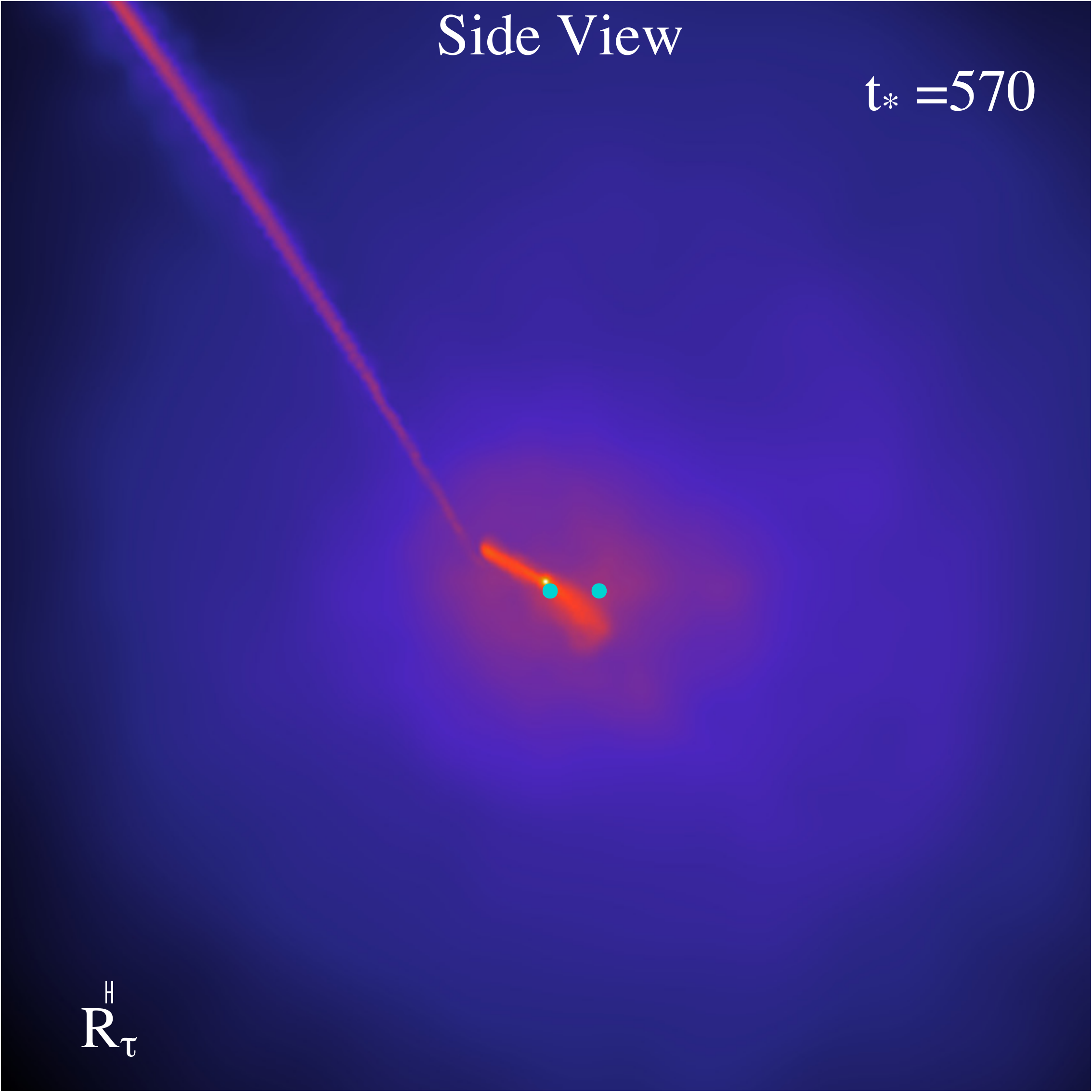}
		\end{minipage}
		\hspace{-0.3cm}	
		\begin{minipage}[c]{0.24\textwidth}
			\includegraphics[width=\textwidth]{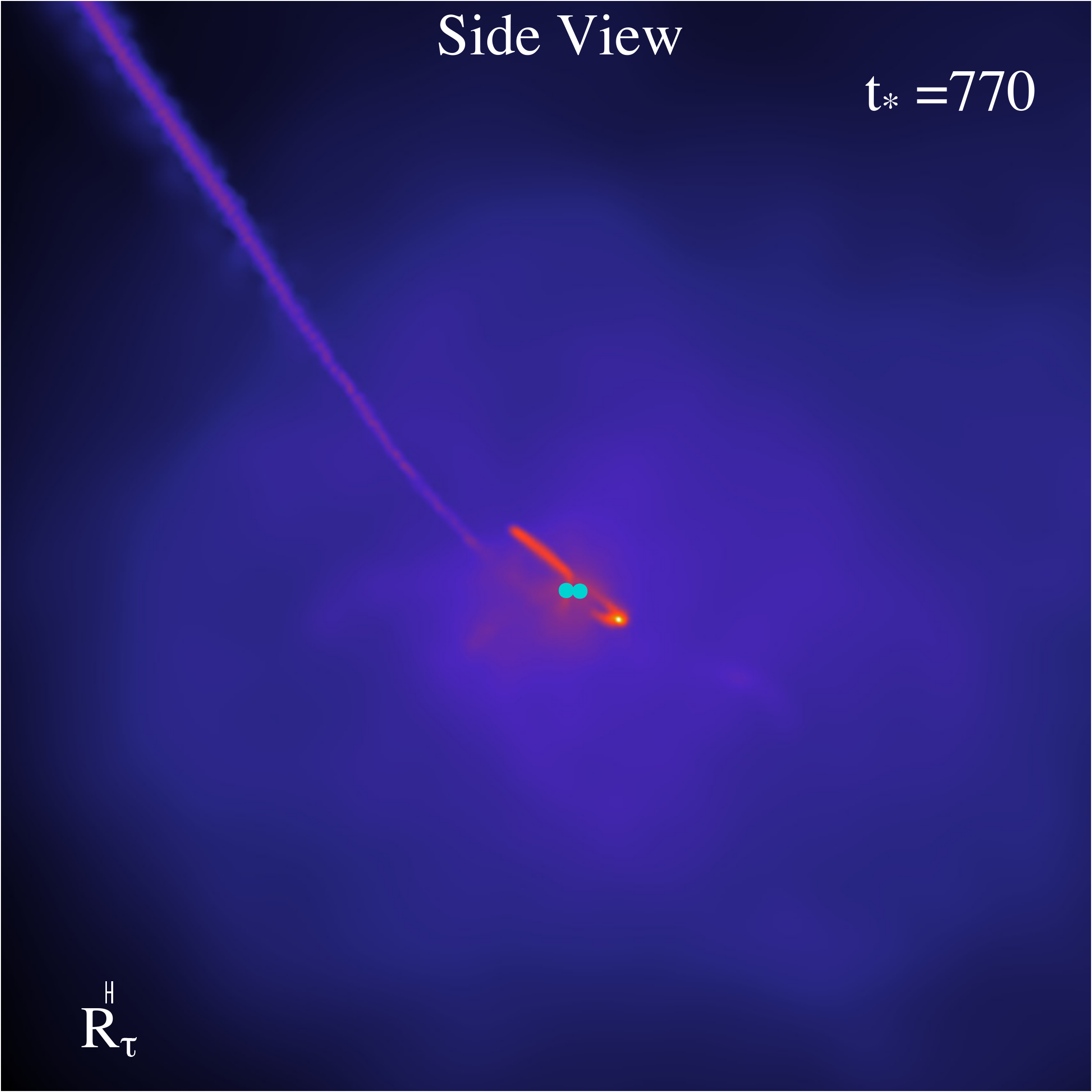}
		\end{minipage}
		\hspace{0.3cm}
		\begin{minipage}[c]{0.24\textwidth}
			\includegraphics[width=\textwidth]{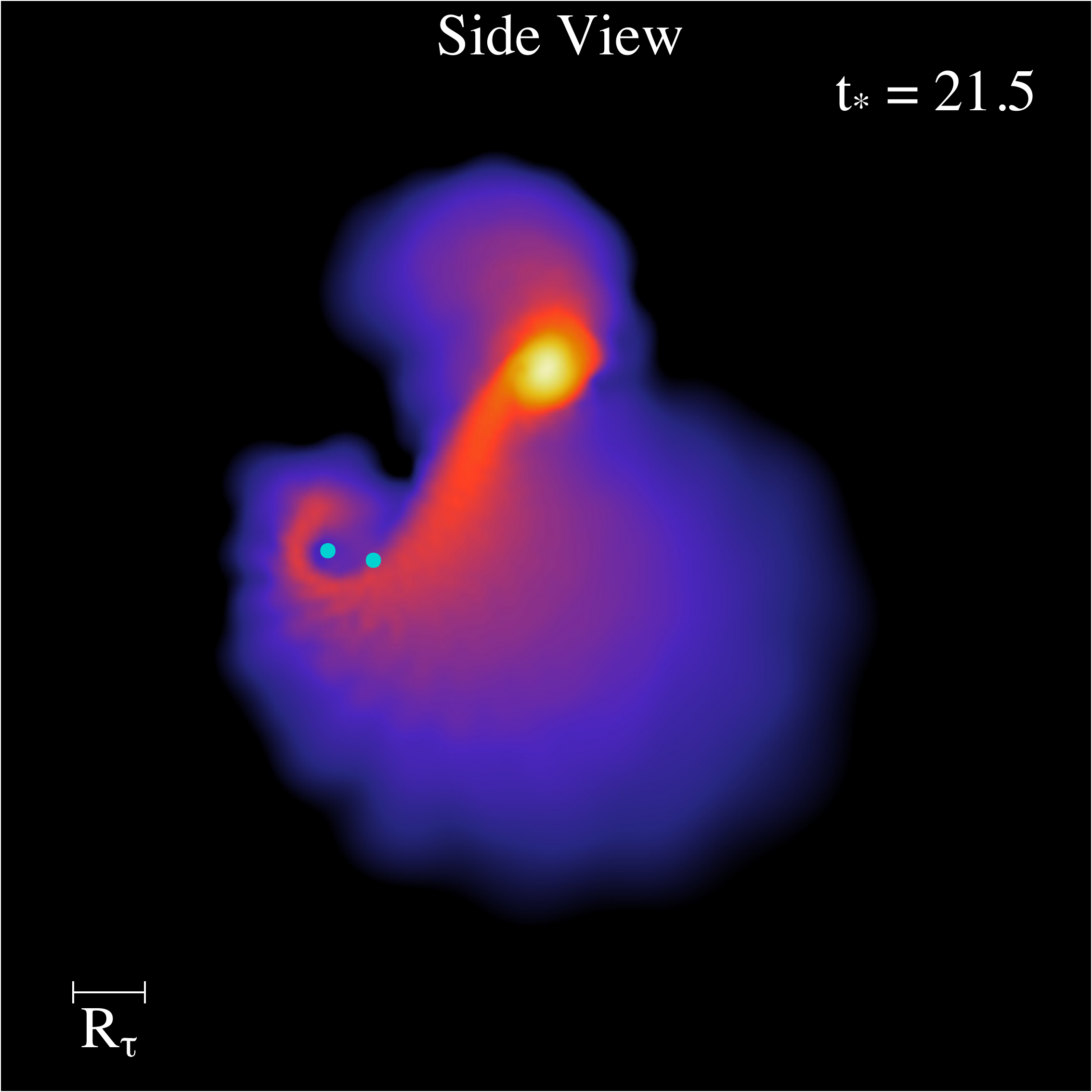}
		\end{minipage}
		\hspace{-0.3cm}
		\begin{minipage}[c]{0.24\textwidth}
			\includegraphics[width=\textwidth]{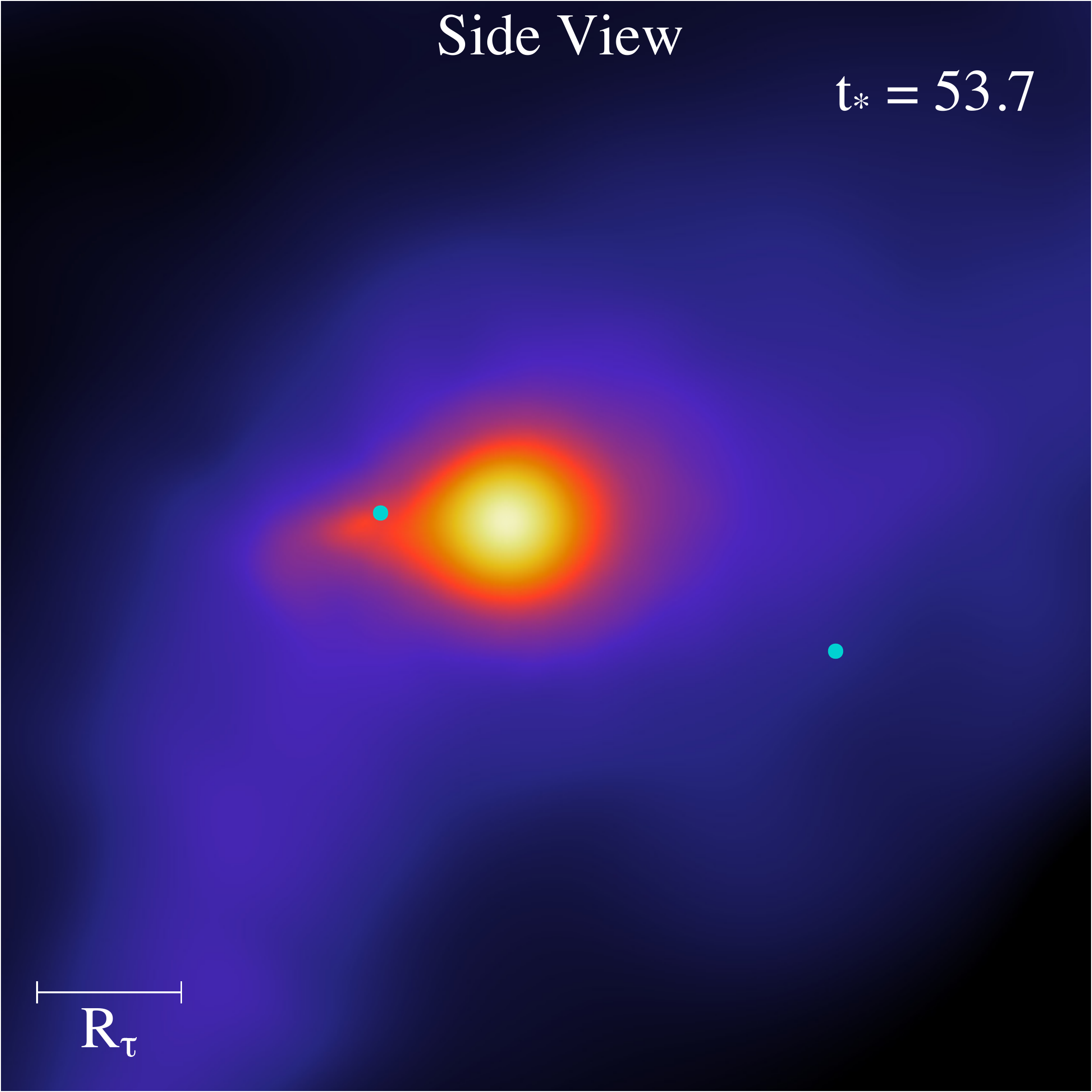}
		\end{minipage}
	\end{center}
	
	\caption{A comparison between  two OS simulations. Here $t_{*}$ denotes the time in units of the star's dynamical timescale while $R_{\tau}$ shows the scale of the individual BH tidal radius. {\it Left Panel:}  Here we show the simulation snapshots for the OS shown in Figure \ref{fig:unique}.  The additional side views  plotted here clearly show how the orientation  of the accretion disk changes between the multiple disruptions. {\it Right Panel:}  Shown are the simulation snapshots for the MOS discussed in Section~\ref{subsubsec:MOS}.  The  orbital view shows the  two interactions  that take place until  full disruption of the star. In the side view snapshots  one can clearly  see that the the orbital angular momentum of the binary is altered by the 3-body interaction. This change is significant in this case due  to the higher mass ratio between the star and the BBH. The simulation parameters for the MOS  are: $N=10^5$, $\mstar = 5 \ \Msun$, $\rstar = 6 \ \Rsun$, $\Gamma = 4/3$, $\mbh{1} = \mbh{2} = 10 \ \Msun$, $d = 21.49 \ \Rsun$, $v_{\infty} = 30 \rm{\ km/s}$, $e = 0.5$.}
	\label{fig:OSMOSSIM}
	\vspace{0.2cm}
\end{figure*}

\subsection{Simulation Results}
\label{subsec:results}
In Section \ref{subsec:BBHdyn} we have outlined three representative scenarios for LBBH TDEs: SS, CS and OS.  In the SS case we have ${\rtau} \ll d$ and $\aninety < \rl$ and the event  resembles that  from a single BH TDE in which only one BH accretes. In the CS case we have  ${\rtau} > d$ and the LBBH ends up being embedded in a circumbinary disk.  In the OS case we have ${\rtau} \lesssim d$ and $\aninety > \rl$ and the accretion of the disrupted debris  by both BHs is able to produce multiple TDEs. The simulation results for the various cases outlined here are presented in  Sections \ref{subsubsec:SS}-\ref{subsubsec:MOS} and shown in Figure \ref{fig:unique} and Figure \ref{fig:OSMOSSIM}. 

\subsubsection{The Single Scenario}
\label{subsubsec:SS}

The SS simulation is characterized here by ${\rtau}/d = 0.006$ and $\aninety/\rl = 0.54$. For these ICs,  almost no significant interaction of the disrupted material is expected to occur with the non-disrupting BH. The SS simulation shown here is consistent with the scenario shown in Figure \ref{fig:stats} for an unbound stellar orbit. The {\it top panels} in Figure \ref{fig:unique} show the gas column density in the orbital plane at three different times, which are shown  in  units of the dynamical timescale  of the star. The bound material is observed to  circularize promptly and, as a result, the mass accretion rate is observed to follow the standard mass fallback rate. However, given that  $q=0.066$,  the early shape of the  mass accretion rate curve differs from that derived  by  \citet{2013ApJ...767...25G}, which was calculated assuming $q\ll 1$. By the end of the simulation, the disrupting BH accreted a total mass of $0.1\Msun$ and has an accretion disk with a leftover mass of about $0.12\Msun$ and whose angular momentum $\jdisk$ is inclined about $1.75 \ \rmn{rad}$ with respect to the orbital angular momentum of the binary $\jbin$. This angle is consistent with that of  the star's angular momentum at the moment of disruption. Assuming that the bound $\approx 0.12 \ \Msun$ of material  is accreted by the BH, the resultant spin magnitude will be $ S_{1}\approx 0.05$, resulting in an anti-aligned effective spin of $\ceff \approx -0.006$.  

\begin{figure*}[t]
	\begin{center}
		\begin{minipage}[c]{0.48\textwidth}
			\includegraphics[width=\textwidth]{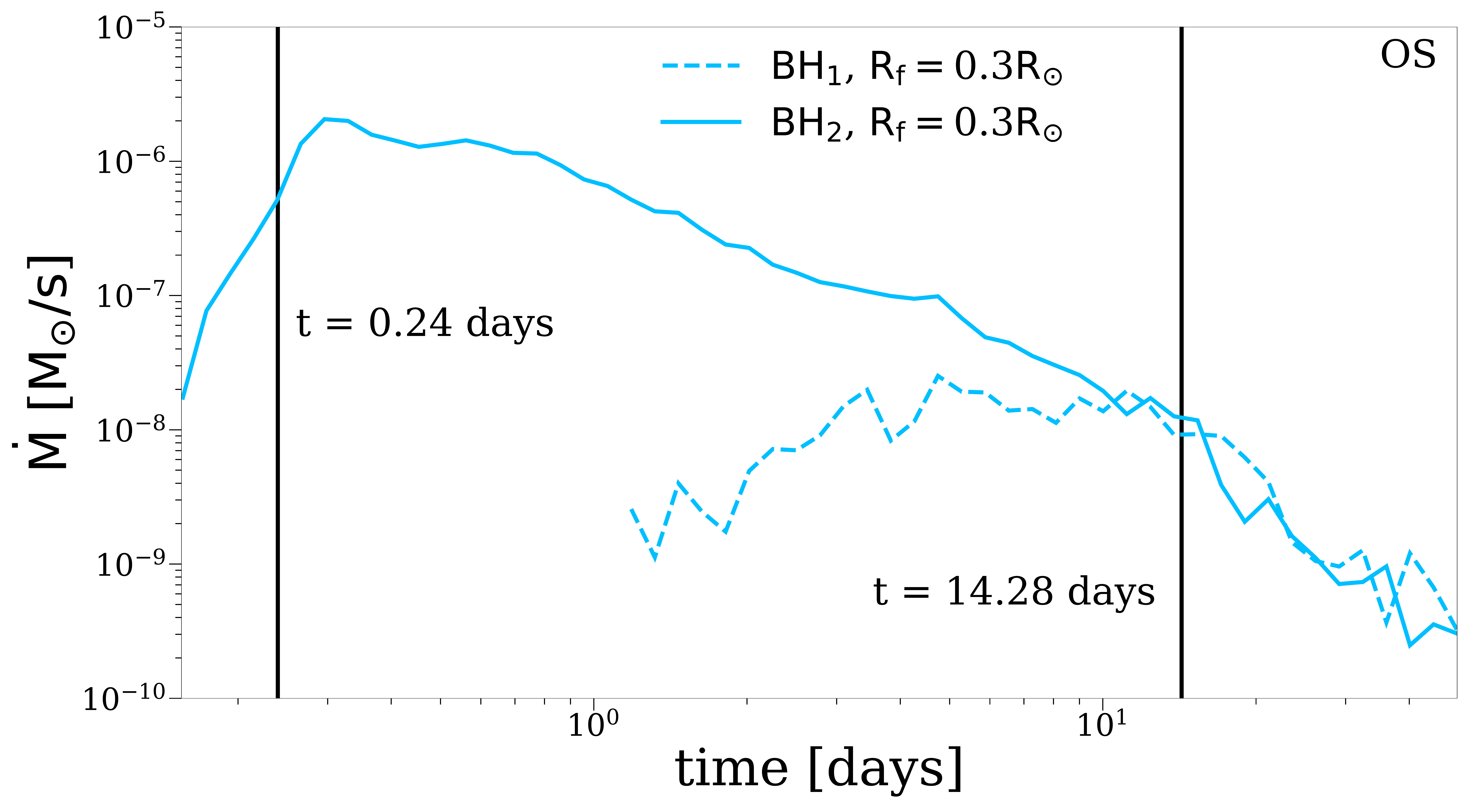}
		\end{minipage}
		\hspace{0.4cm}
		\begin{minipage}[c]{0.48\textwidth}
			\includegraphics[width=\textwidth]{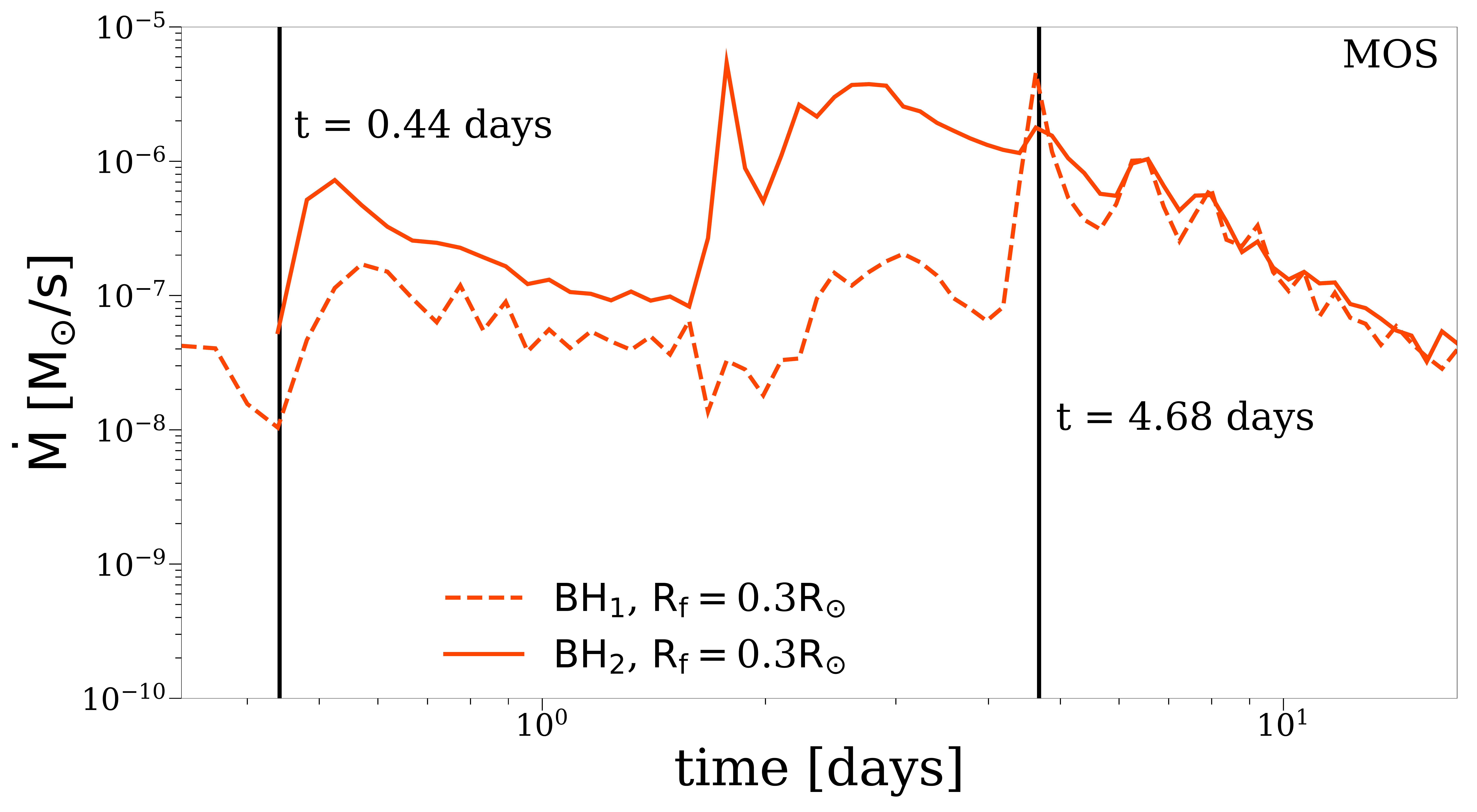}
		\end{minipage}
	\end{center}
	
	\vspace{-0.5cm}
	
	\begin{center}
		\hspace{0.75cm}
		\begin{minipage}[c]{0.22\textwidth}
			\includegraphics[width=\textwidth]{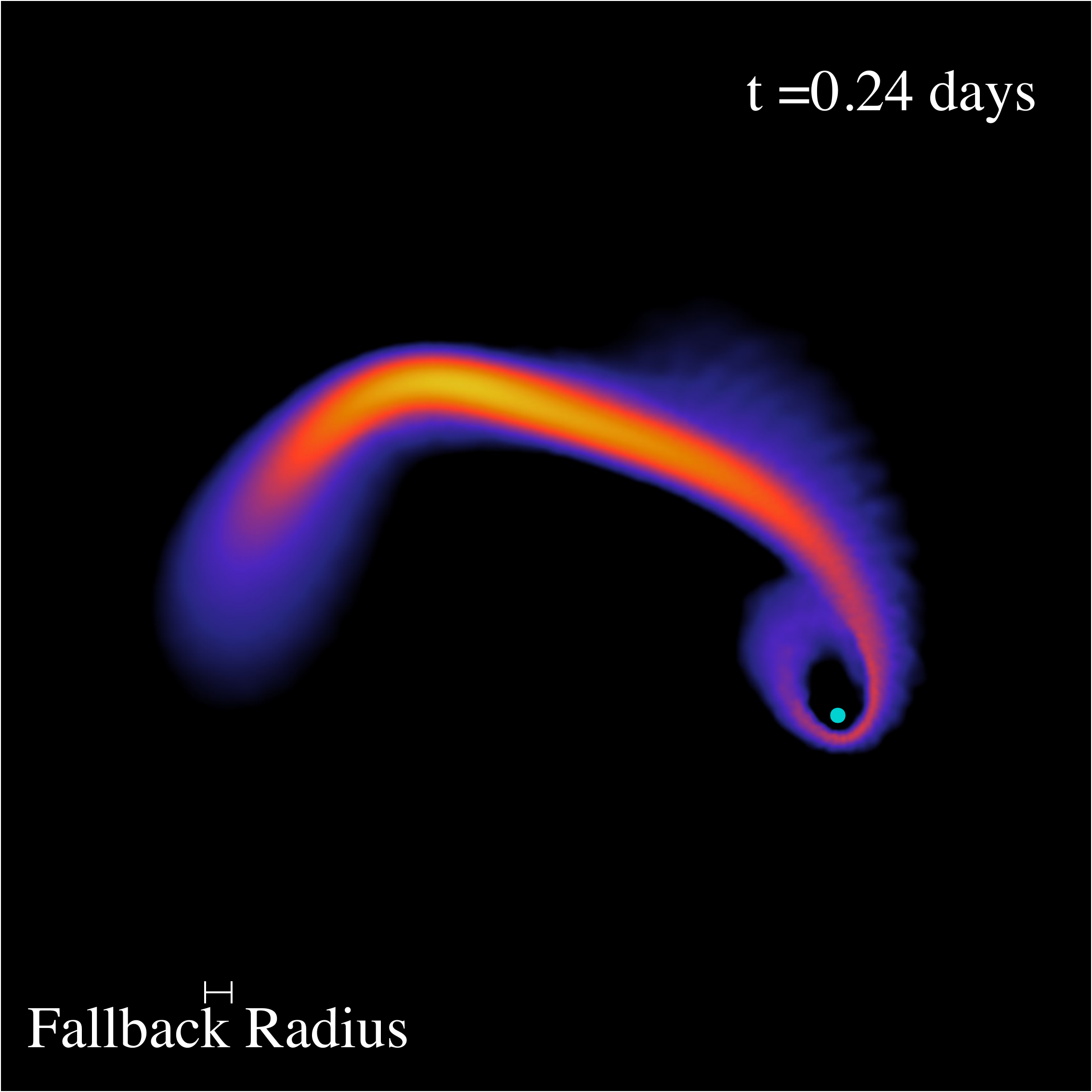}
		\end{minipage}
		\hspace{-0.25cm}
		\begin{minipage}[c]{0.22\textwidth}
			\includegraphics[width=\textwidth]{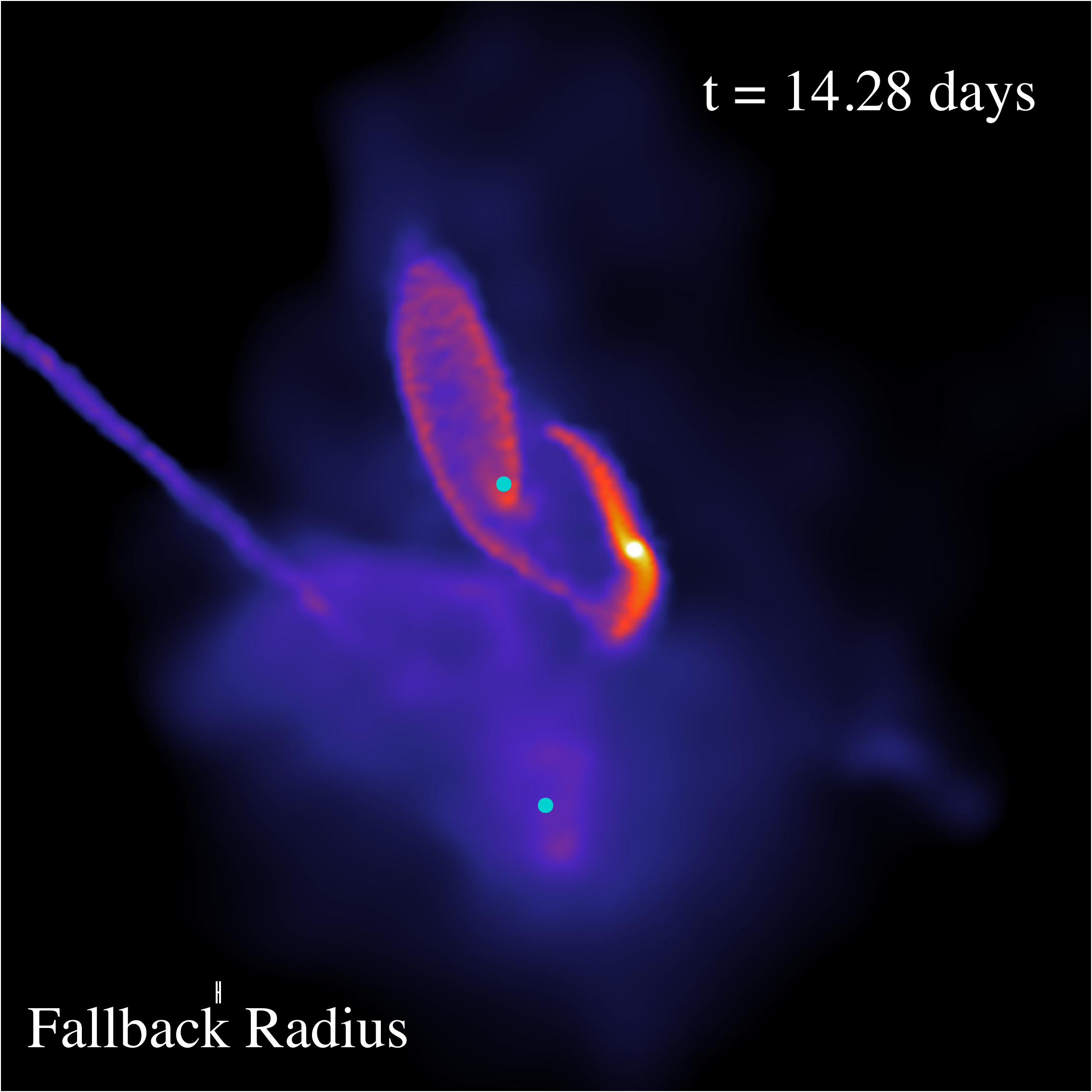}
		\end{minipage}
		\hspace{1.15cm}
		\begin{minipage}[c]{0.22\textwidth}
			\includegraphics[width=\textwidth]{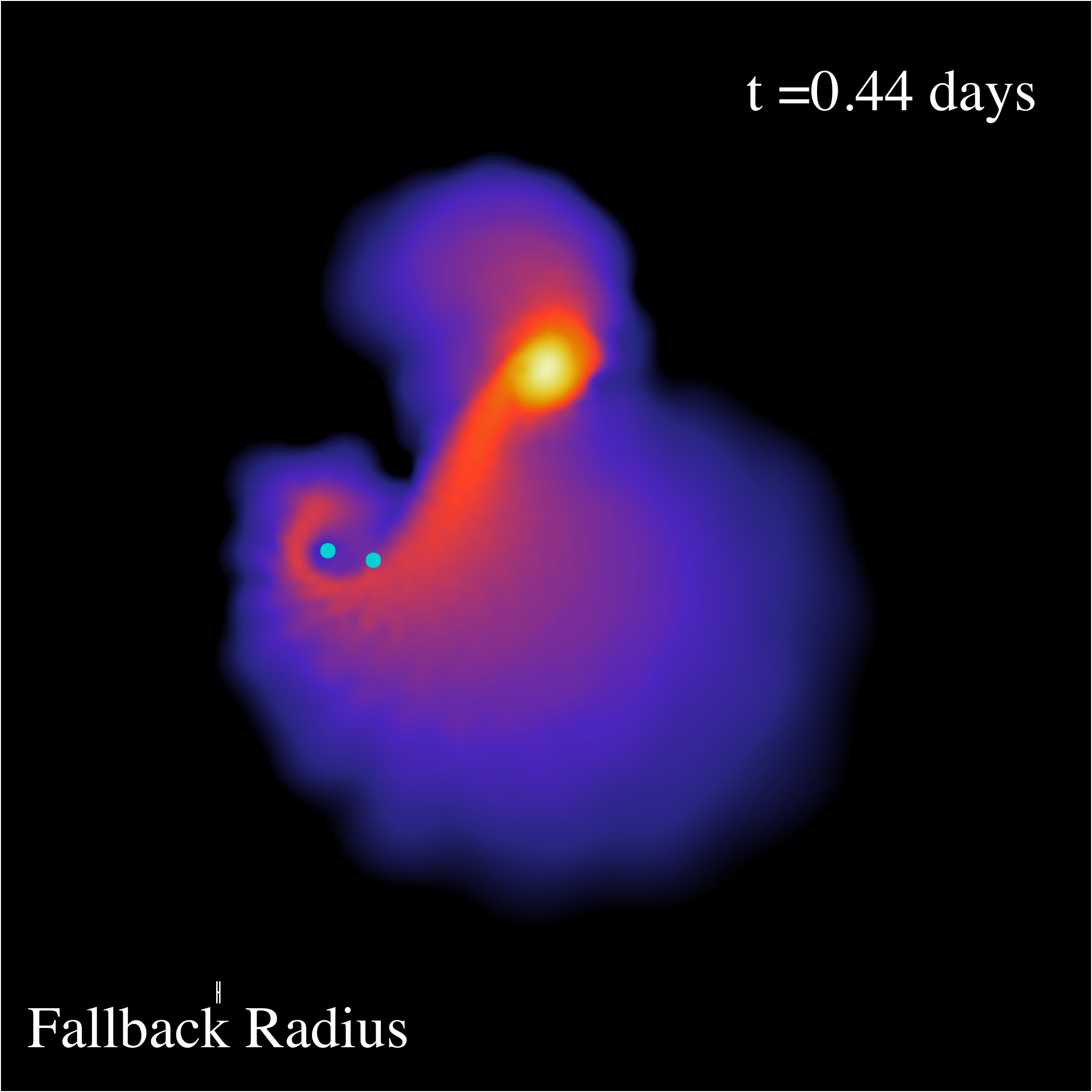}
		\end{minipage}
		\hspace{-0.25cm}	
		\begin{minipage}[c]{0.22\textwidth}
			\includegraphics[width=\textwidth]{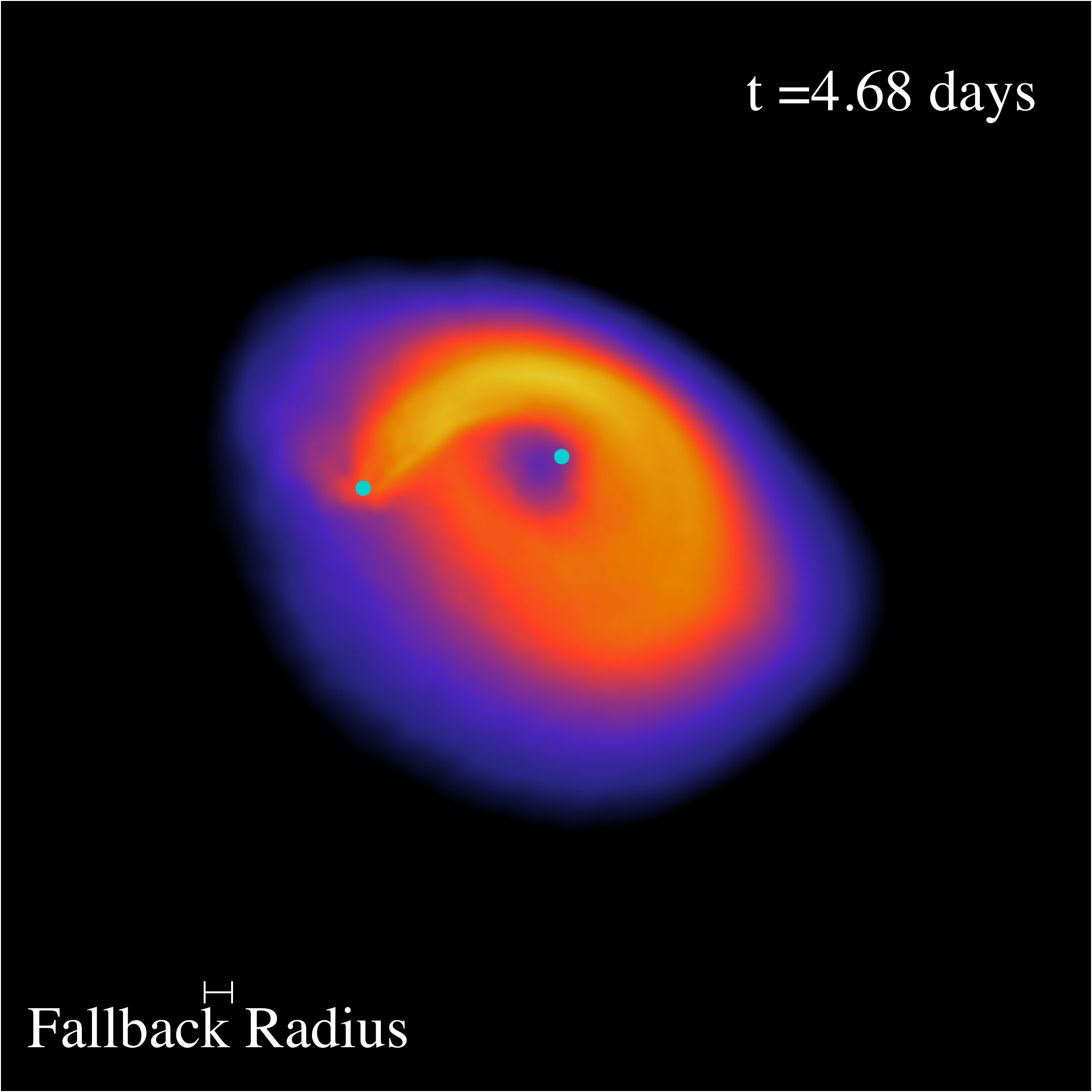}
		\end{minipage}
	\end{center}
	
	\caption{The mass accretion histories experienced by both BHs in the OS and MOS. Here  $\rm{R_{\rm f}}$ is the {\it Fallback Radius}  of the surface at which the mass flux is calculated. All snapshot times are measured from the time  the most bound material is accreted. {\it Left Panel:} The top panel shows accretion curves for both BHs for the OS.  A total of four tidal interactions take place. The first is responsible  for feeding significant mass to the disrupting, while the non-disrupting BH accretes material as it plunges into the disk around the disrupting BH. The bottom panels show snapshots of the simulation at 0.24 and 14.3 days (shown as vertical lines in the  top panel). {\it Right Panel:}The top panel shows the rate of mass accretion onto both BHs. In contrast to the OS, both BHs accrete promptly due to the larger amount of stellar material available. After the second disruption the accretion  curves are almost identical. In this case,  both BHs to accrete a notable percentage of their own mass during the disruption. The bottom panels show snapshots of the simulation at two specific times: 0.16 and 4.42 days (shown as vertical lines in the  {\it Top Panel}).} \label{fig:OSMOSDIFF}
\end{figure*}

\subsubsection{The Circumbinary Scenario}
\label{subsubsec:CS}
The CS simulation is parametrized by ${\rtau}/d = 2.47$. The tidal radii of each BH overlap and encompass the binary, resulting in a disruption where bound material forms a circumbinary disk. At the moment of disruption the orientation of the angular momentum of the star's CM with respect to $\jbin$ is approximately $2.44 \ \rmn{rad}$. As the most bound material  returns to pericenter the binary exerts a torque on the stream and, as a result,  alters  the angle of $\jdisk$ to $\approx 2 \ \rmn{rad}$; see Section \ref{subsec:evolution} and Figure \ref{fig:disks}. The disk rapidly circularizes due to hydrodynamical dissipation at pericenter as well as collisions between the returning stream caused by the time changing binary potential ({\it middle panels} in Figure \ref{fig:unique}). The material residing in the disk is slowly accreted onto both BHs through viscous dissipation. We stopped the simulation at approximately ten percent of the time it would take to ingest the entire disk and found that each BH accreted about $0.01\Msun$ and the accretion disk has  $~0.3 \Msun$ of gas leftover. If we assume that this material is evenly accreted by both BHs, the resultant spin magnitudes will be $S_{1}=S_{2} \approx 0.036$ and, given that the spin angles of each BH are aligned with $\jdisk$, $\ceff \approx- 0.015$.

\subsubsection{The Overflow Scenario}
\label{subsubsec:OS}
\begin{figure*}[t]
	\begin{center}
		\begin{minipage}[c]{0.3\textwidth}
			\includegraphics[width=\textwidth]{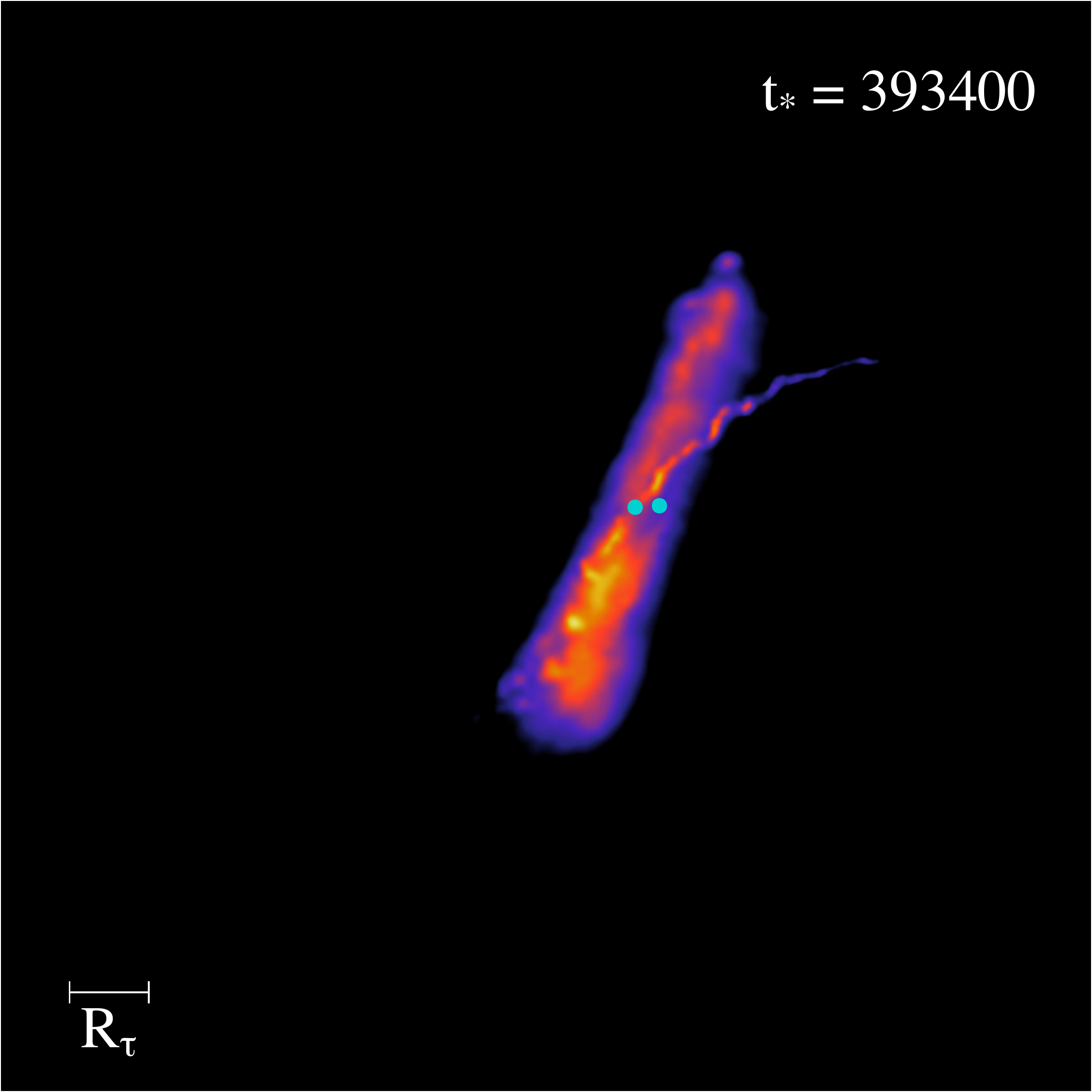}
		\end{minipage}
		\hspace{-0.22cm}
		\begin{minipage}[c]{0.3\textwidth}
			\includegraphics[width=\textwidth]{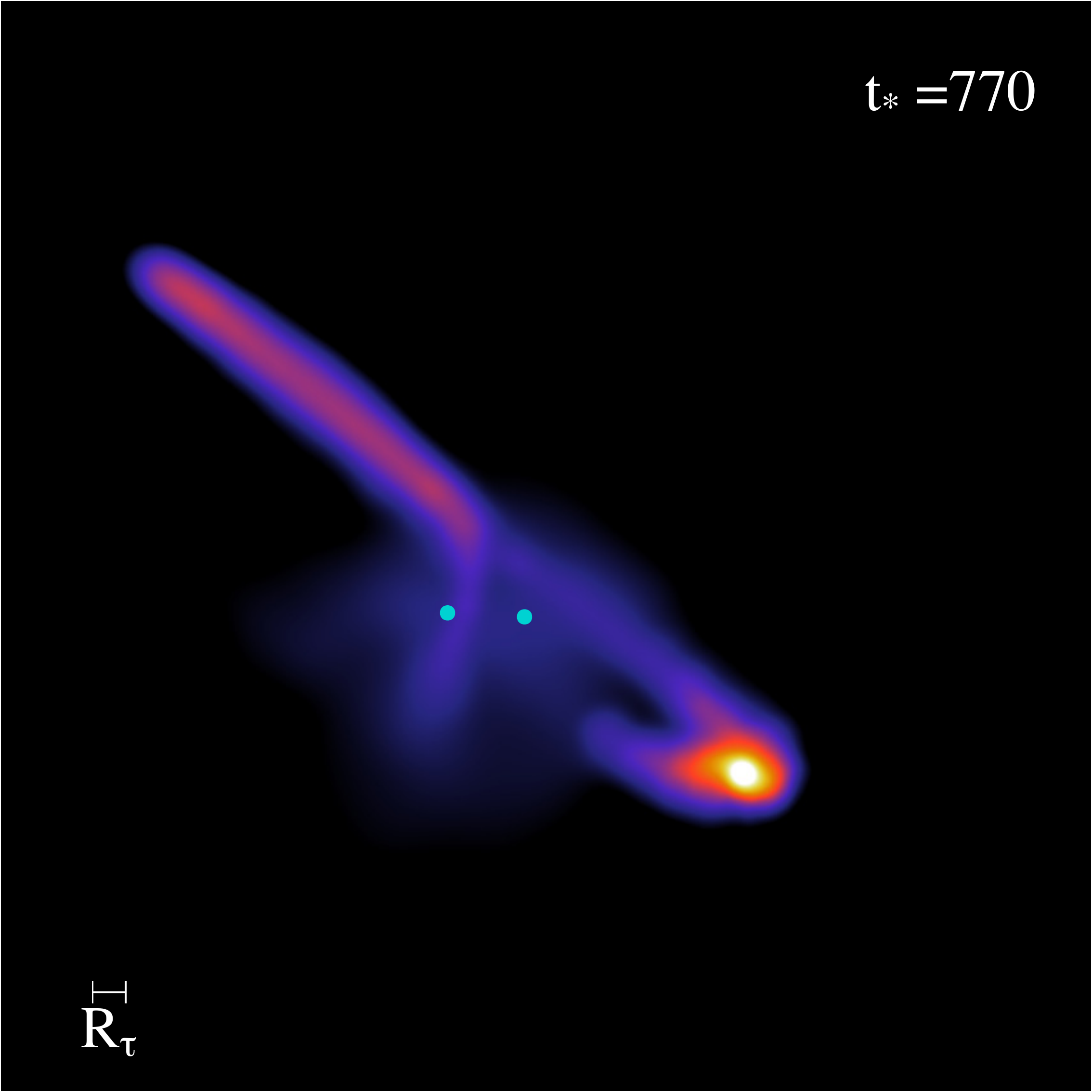}
		\end{minipage}
		\hspace{-0.22cm}	
		\begin{minipage}[c]{0.3\textwidth}
			\includegraphics[width=\textwidth]{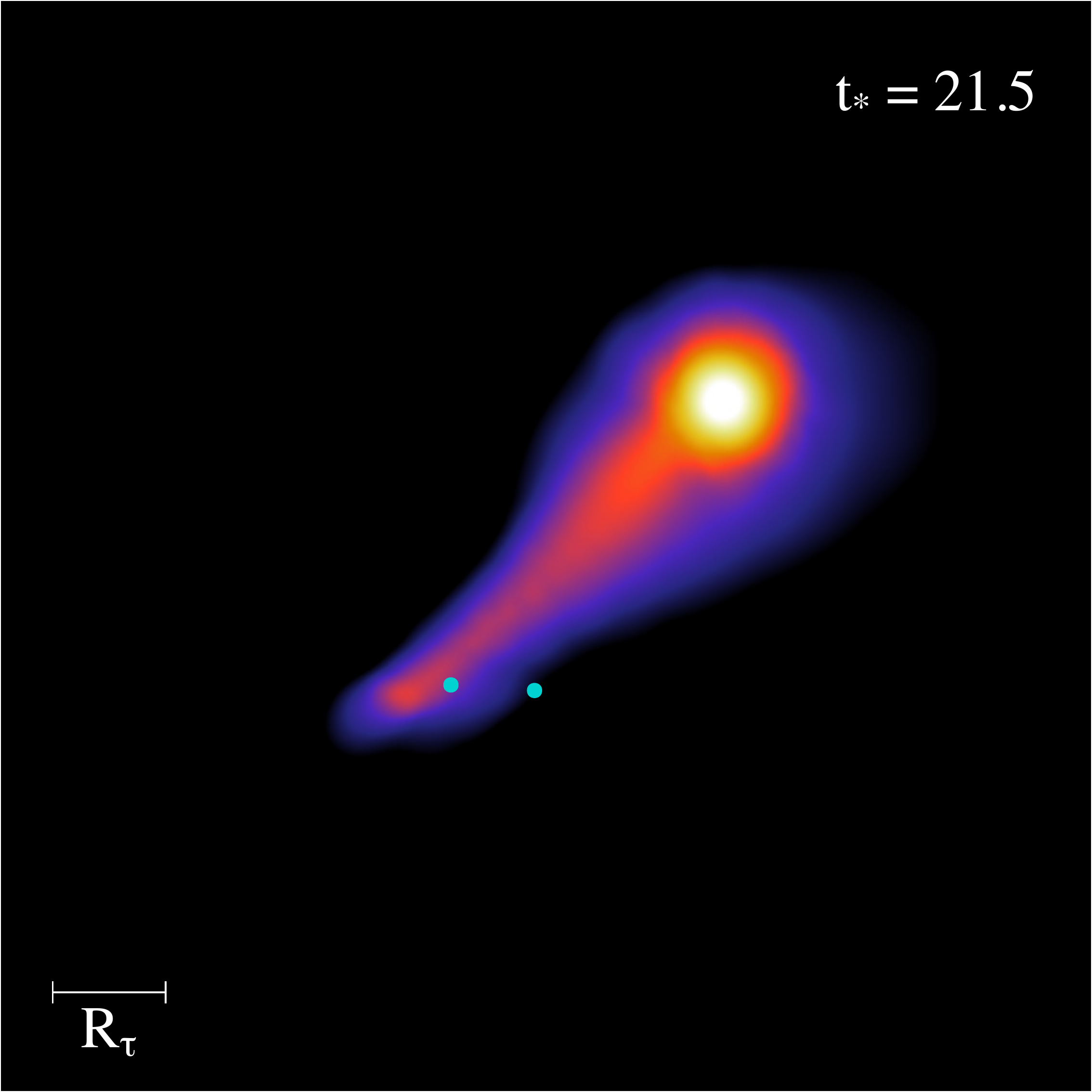}
		\end{minipage}
	\end{center}
	
	\caption{The structure of the accretion disks  formed during the circumbinary, overflow, and massive overflow scenarios. Here $t_{*}$ and $R_{\tau}$ denote the time in units of the star's dynamical timescale and the individual tidal radius for each panel respectively. {\it Left Panel:} Snapshot showing the accretion disk structure  at the end of the circumbinary scenario (CS) simulation. The angle of the angular momentum of the disk, $\jdisk$, relative to the orbital angular momentum of the binary, $\jbin$, is about $2.0 \ \rmn{rad}$. {\it Middle Panel:} Snapshot showing the accretion disk structure after the third TDE (out of a total of four before full disruption)  in the overflow scenario (OS). The angle of $\jdisk$ relative to $\jbin$ is approximately $0.69 \ \rmn{rad}$. {\it Right Panel:} Snapshot of the accretion disk after the initial TDE in the massive overflow scenario (MOS) case. The angle of $\jdisk$ relative to $\jbin$ is $\approx 0.85 \ \rmn{rad}$.}
	\label{fig:disks}
\end{figure*}

The OS simulation is characterized here by  ${\rtau}/d = 0.06$ and $\aninety/\rl = 5.44$. This guarantees that after the disruption, a significant amount of bound disrupted material will be able to reach the sphere of influence of the non-disrupting BH. Within this scenario, accretion onto both BHs can occur, which might result in temporary BH spin alignment or anti-alignment. The star survives after the initial disruption leading to multiple resonant TDEs, as can be seen in the {\it bottom panels} of  Figure \ref{fig:unique}. A total of four interactions take place with the same BH in this scenario until the star is fully disrupted. The angular momentum of the star with respect to $\jbin$ changes in  each disruption. By the end of the simulation, the mass accreted by the disrupting and non-disrupting BHs is $0.19 \ \Msun$ and $0.02 \ \Msun$, respectively. The resultant angles  are $1.58 \ \rmn{rad}$ and $0.24 \ \rmn{rad}$ with respect to  $\jbin$ for the disrupting and non-disrupting BHs, respectively. The first disruption provides the majority of the accreted mass for the disrupting BH, while the the non-disrupting BH accretes mass as it returns to the pericenter of the binary orbit. Therefore, the angle for the disrupting BH is similar to that of the star's angular momentum with respect to $\jbin$ at the time of the first disruption, while the non-disrupting BH's angle is aligned with $\jbin$; see Section \ref{subsec:evolution} and Figure~\ref{fig:disks}. We obtain $S_1 \approx 0.04$  and $S_2 \approx 0.006$ which leads to a final  $\ceff \approx 0.003$.

\subsubsection{The Massive Overflow Scenario}
\label{subsubsec:MOS}
The changes in spin magnitude obtained in the scenarios  discussed previously are  expected to be small  given that $S_{\rmn{max}}\left(q = 0.067\right) = 0.12$.  More sizable changes are expected for larger values of $q$. Motivated by this,  we run a simulation in which $q=0.5$, which we refer to as the \textit{massive overflow scenario} (MOS). The MOS simulation is characterized by ${\rtau}/d = 0.35$ and $\aninety/\rl = 16.14$. A comparison between the OS and MOS is shown  in Figure \ref{fig:OSMOSSIM}.

Both OS and MOS simulations lead to multiple disruptions and result in accretion onto both BHs. However, the $\mdot$ curves shown in Figure \ref{fig:OSMOSDIFF} are significantly different. In the OS, accretion onto the disrupting BH proceeds   like in a canonical TDEs, showing a fast rise  and a subsequent power-law decay. Accretion onto the non-disrupting BH, which occurs as it plummets into the accretion disk around the disrupting BH, is observed to be delayed and increases at a slower rate. In the MOS panel, accretion onto both BHs occurs at a similar time and the $\mdot$ curves for both BHs are rather similar yet differ from the  canonical TDEs. In this case the first disruption was weaker and most of the material was made available to the BHs until after the second disruption (Figure \ref{fig:OSMOSDIFF}). The star gets considerably closer to the BH during the second encounter and, as a result, the star is completely disrupted. In what follows we refer to the disrupting BH as the one responsible for the second disruption, which provides the vast majority of the mass supply. The accretion disk that forms  after the second disruption can be seen in the right bottom panel of Figure \ref{fig:OSMOSDIFF} and is observed to be very extended, making it easy for the non-disrupting BH to accrete a substantial amount of material, especially since the binary orbit is highly eccentric and the BH will eventually plunge into the accretion disk.

The mass accreted by the disrupting and non-disrupting BH at the end of the simulation is $0.91 \ \Msun$ and $0.40 \ \Msun$ respectively. This leads to $S_1 \approx 0.283$ at angle $2.2 \ \rmn{rad}$ with respect to $\jbin$ for the disrupting BH and $S_2 \approx 0.136$ at angle $0.14 \ \rmn{rad}$ with respect to $\jbin$ for the non-disrupting BH,  leading to $\ceff \approx -0.019$. The spin angle of the disrupting BH  is  consistent with the angle with respect to $\jbin$ of the star at the time of the second disruption. The non-disrupting black hole accretes the majority of the mass  in the plane of the binary, as in the OS case.  We note that the spin angle in these interactions can change in  due to multiple encounters, as can be clearly  seen in Figure \ref{fig:OSMOSSIM}  for the OS scenario  (see Section \ref{subsec:evolution} for further discussion).

\pagebreak
\section{Discussion}
\label{sec:disc}

The detection of GW150914 and subsequent LBBH merger GW observations have opened up many questions about LBBH formation history. Individual BH spins within the binary are often used to infer  the specific  formation channel. In this paper we have explored the possibility and consequences of a LBBH experiencing a TDE during its lifetime. The accretion that follows from a TDE can possibly spin up each BH and align or anti-align their relative spins. The notion of temporary spin (mis)alignment  contrasts with  the usual assumption that BH spins  are non-evolving and remain unaltered from BH formation to merger. The implications of these tidal interactions are discussed as follows: Section \ref{subsec:evolution} explores spin evolution from single and multiple TDEs; and Section \ref{subsec:transients} presents the possible observational signatures produced by these interactions.

\begin{figure*}[t]
	\begin{center}
		\begin{minipage}[c]{0.49\textwidth}
			\includegraphics[width=\textwidth]{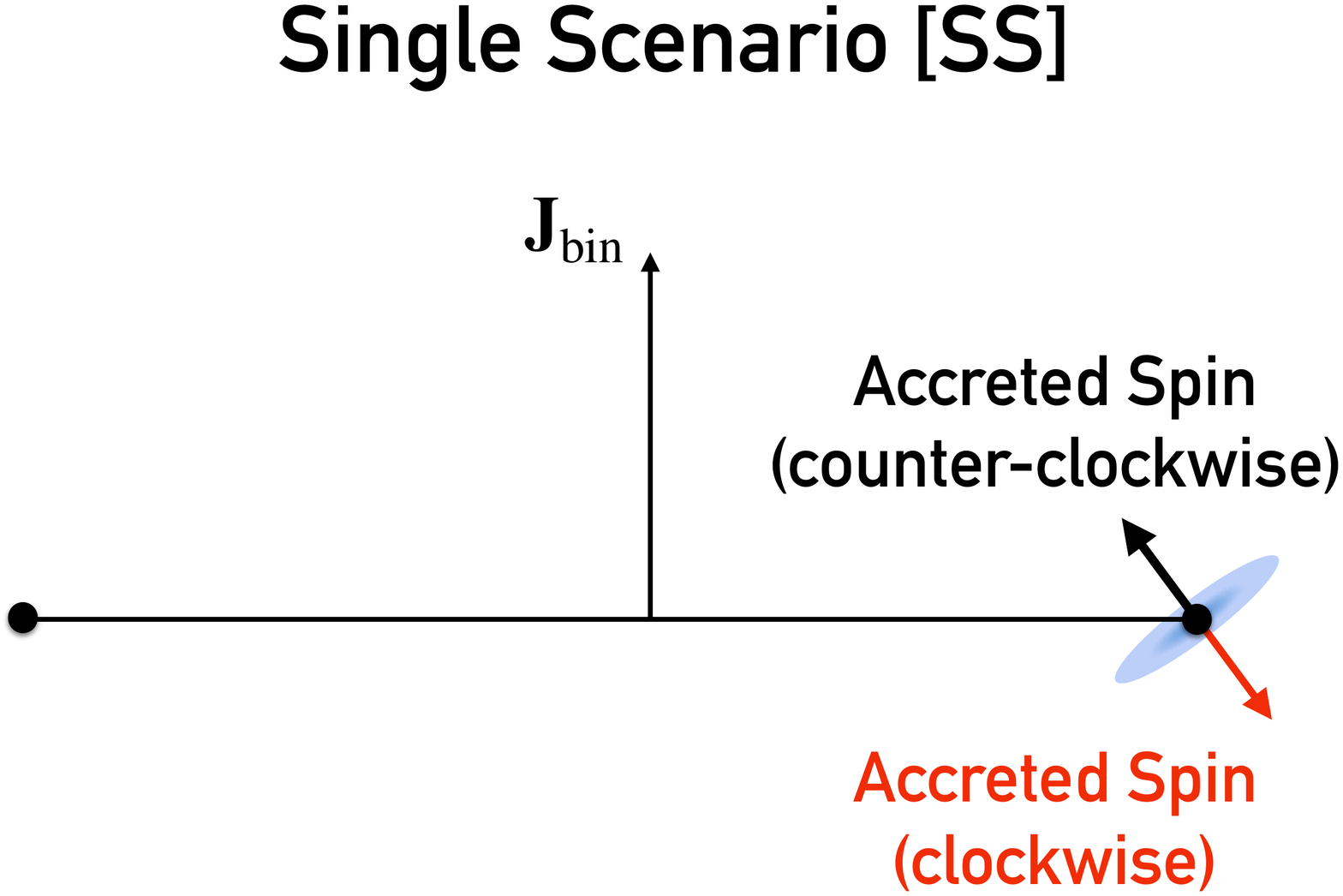}
		\end{minipage}
		\begin{minipage}[c]{0.49\textwidth}
			\includegraphics[width=\textwidth]{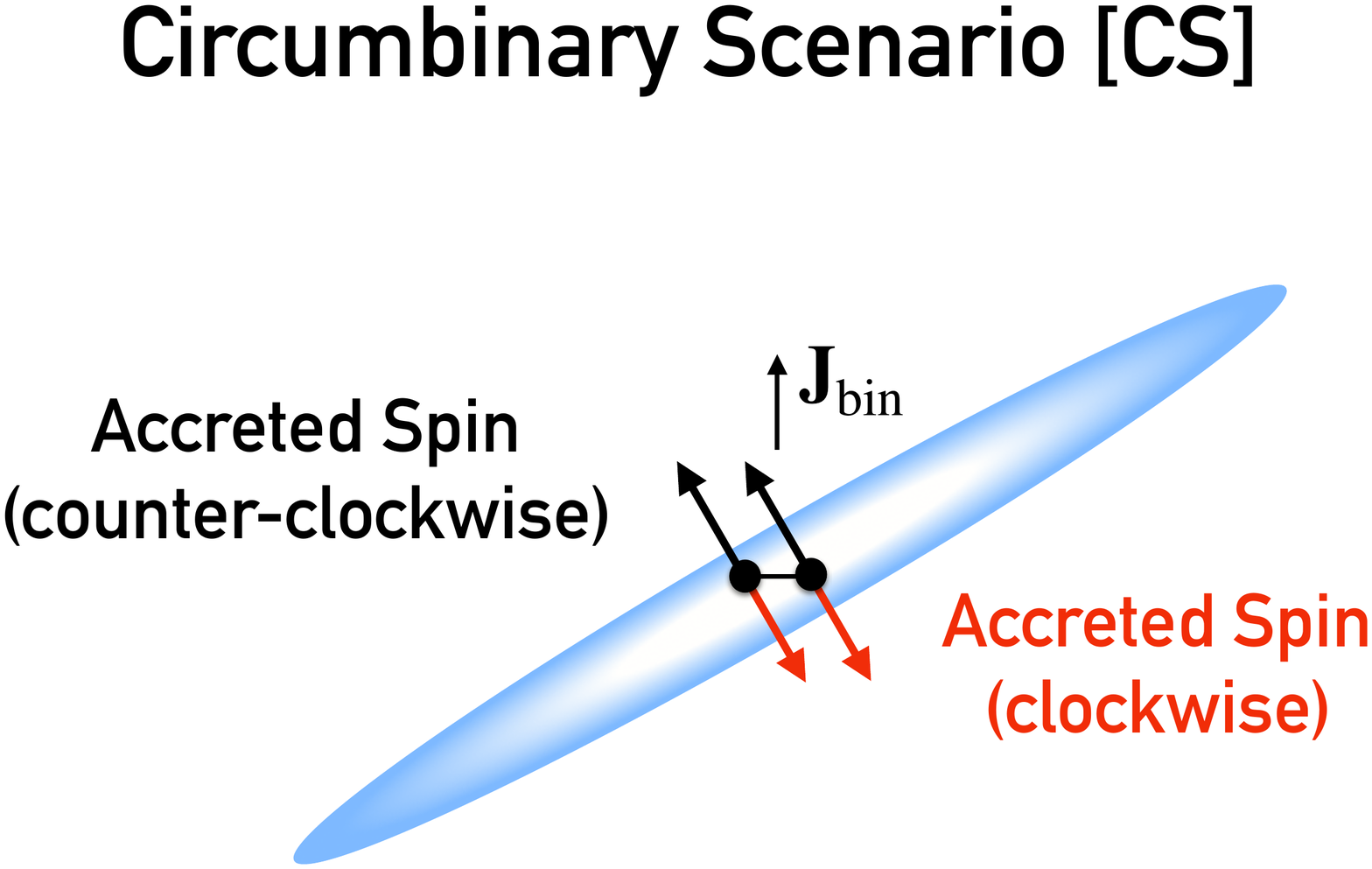}
		\end{minipage}
	\end{center}
	
	\vspace{-1.25cm}
	
	\caption{Diagram illustrating how the accreted spin directions of the  BHs are set by either $\jdisk$ or  $\jbin$. The black and red arrows  show the spin of the BHs expected from accretion of the stellar  debris, whose angular momentum distribution can be  clockwise or counter-clockwise. {\it Left Panel:} In the SS,  a single BH TDE occurs and only the disrupting BH accretes material. The resulting  direction of the BH spin  is expected to be aligned with the angular momentum of the disk $\jdisk$. {\it Right Panel:} In the CS disruption, a circumbinary disk is formed which allows both BHs to accrete material with similar specific angular momentum.}
	\label{fig:SSandCSspins}
\end{figure*}

\subsection{Spin Evolution}
\label{subsec:evolution}

\begin{figure*}
	\begin{center}
		
		\begin{minipage}[c]{0.4\textwidth}
			\includegraphics[width=\textwidth]{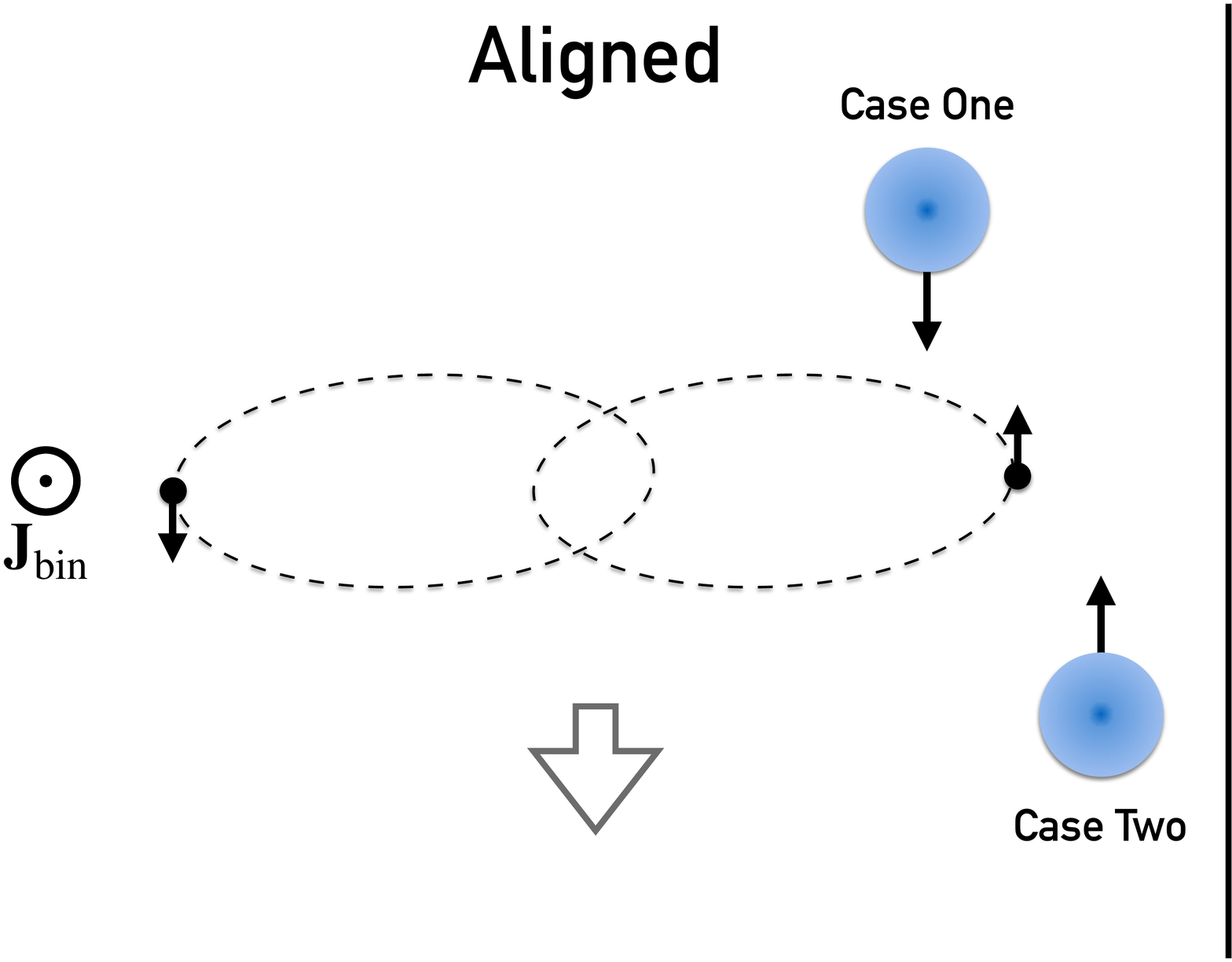}
		\end{minipage}
		\begin{minipage}[c]{0.4\textwidth}
			\includegraphics[width=\textwidth]{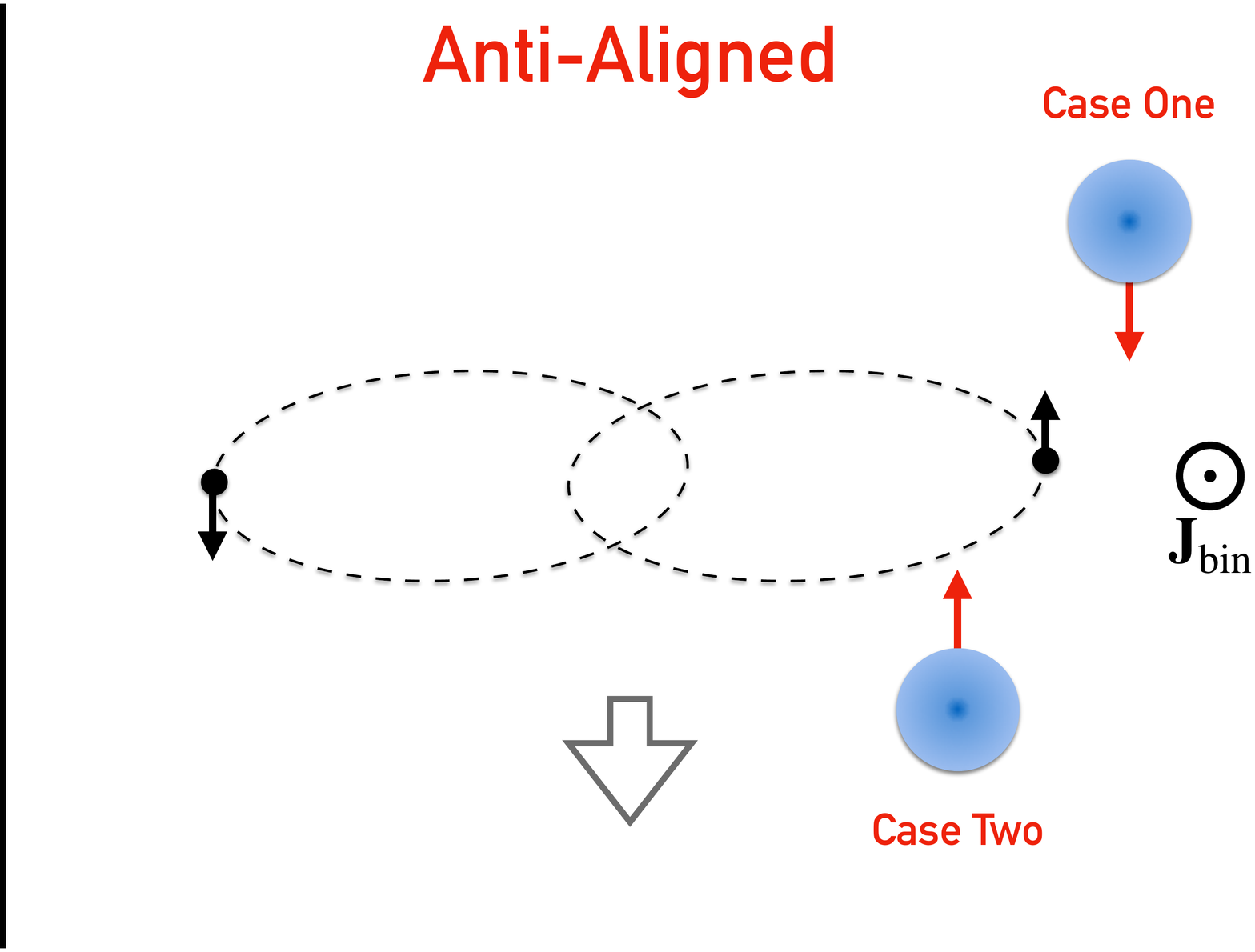}
		\end{minipage}
		
		\vspace{-0.5cm}
		
		\begin{minipage}[c]{0.4\textwidth}
			\includegraphics[width=\textwidth]{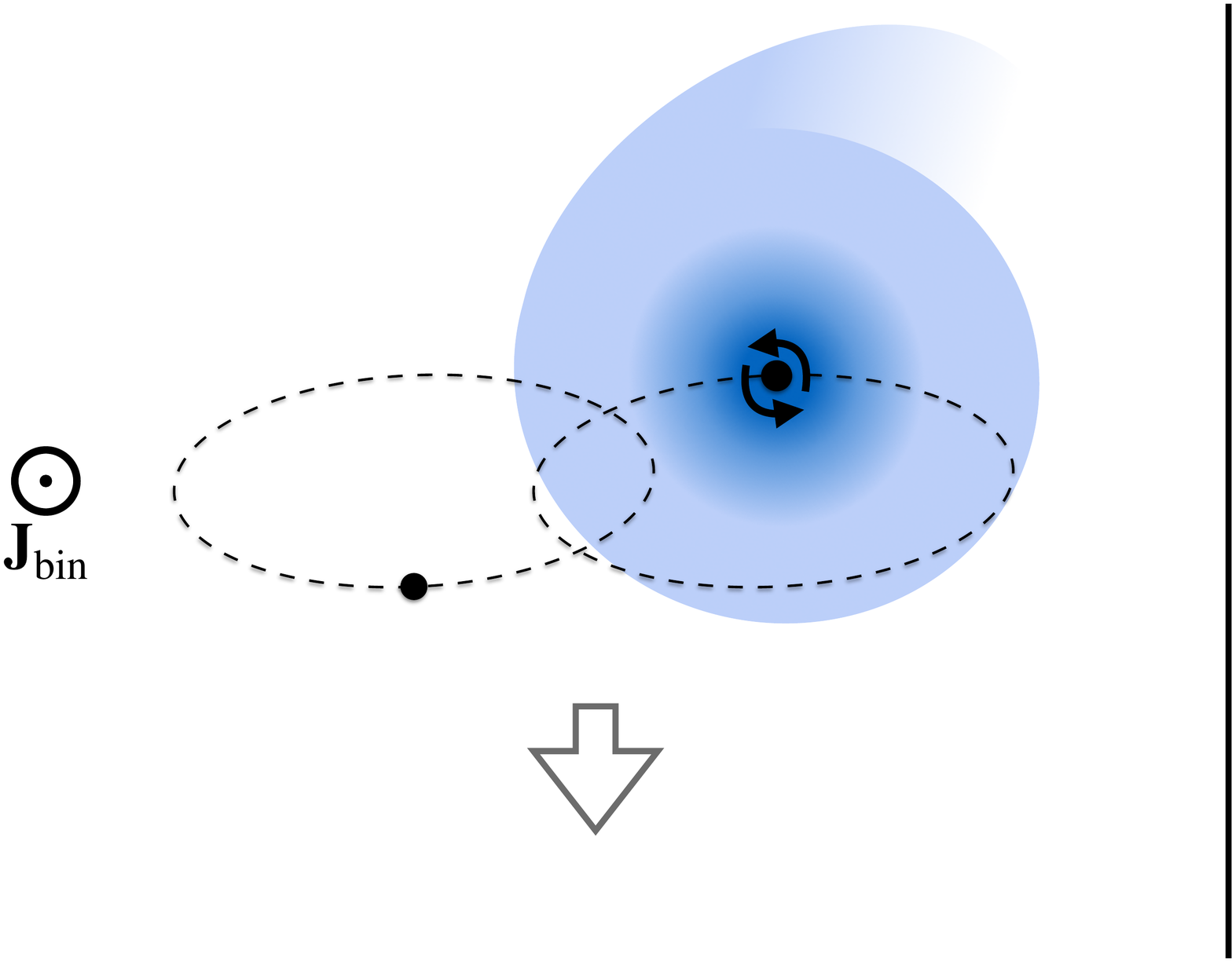}
		\end{minipage}
		\begin{minipage}[c]{0.4\textwidth}
			\includegraphics[width=\textwidth]{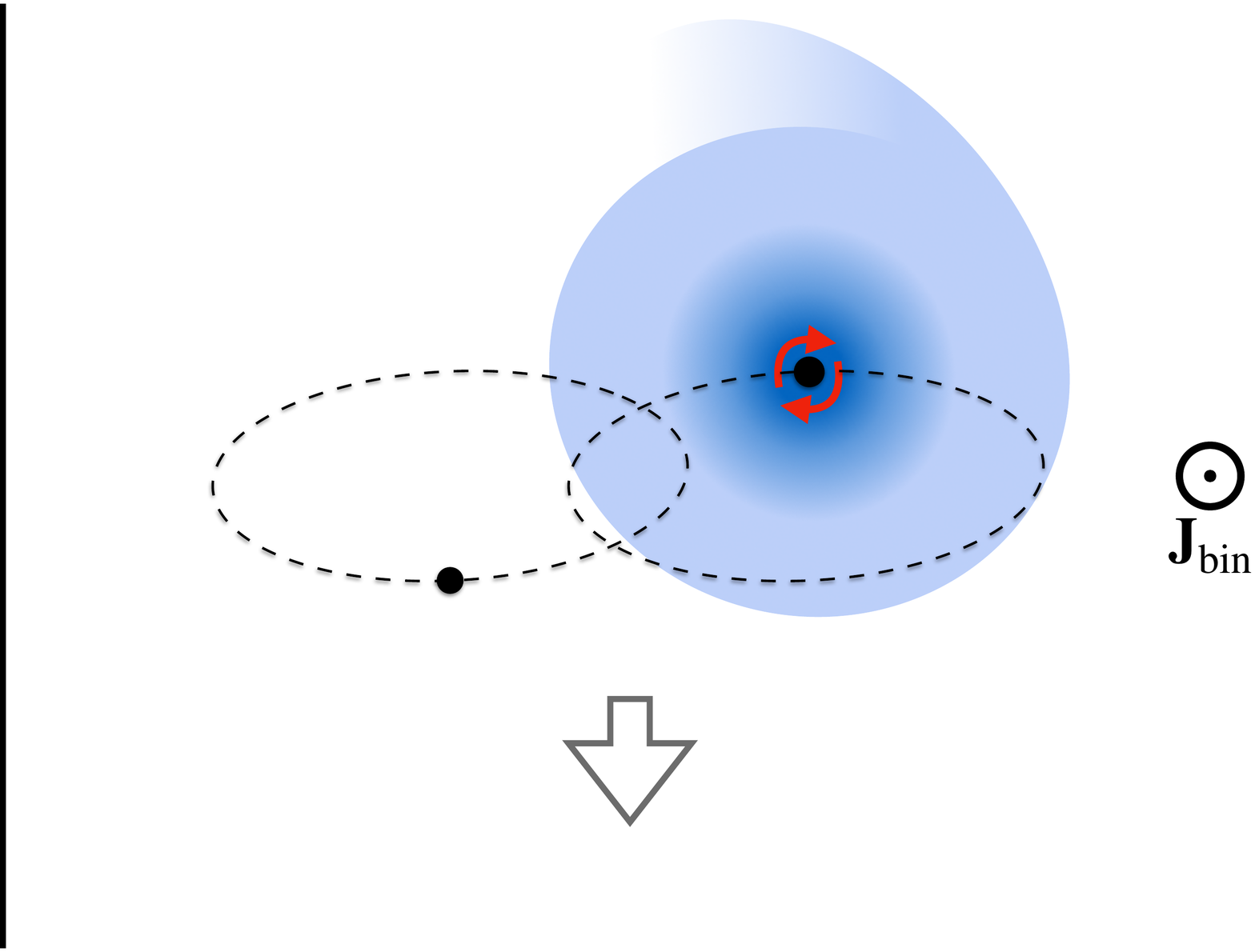}
		\end{minipage}
		
		\vspace{-0.9cm}
		
		\begin{minipage}[c]{0.4\textwidth}
			\includegraphics[width=\textwidth]{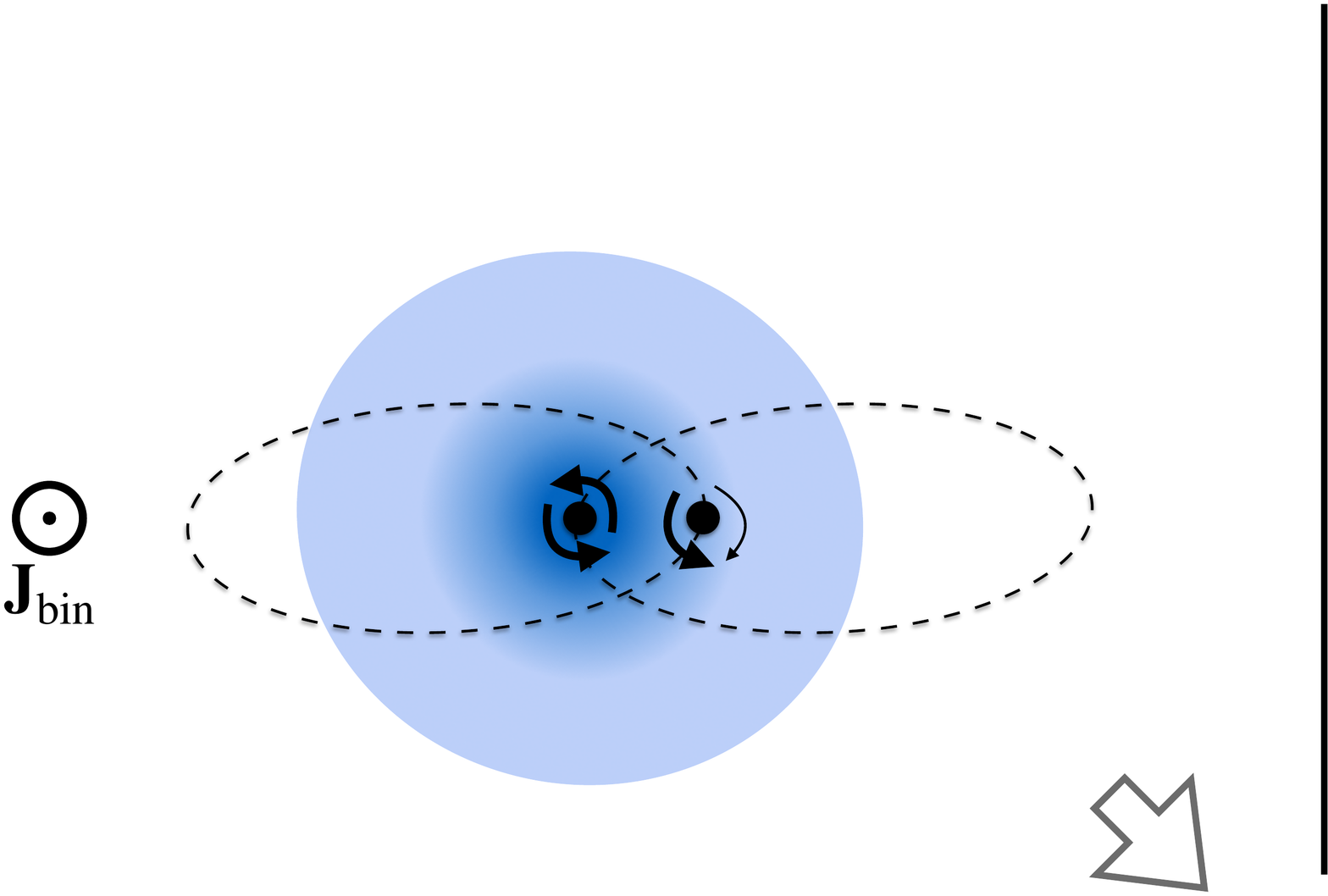}
		\end{minipage}
		\begin{minipage}[c]{0.4\textwidth}
			\includegraphics[width=\textwidth]{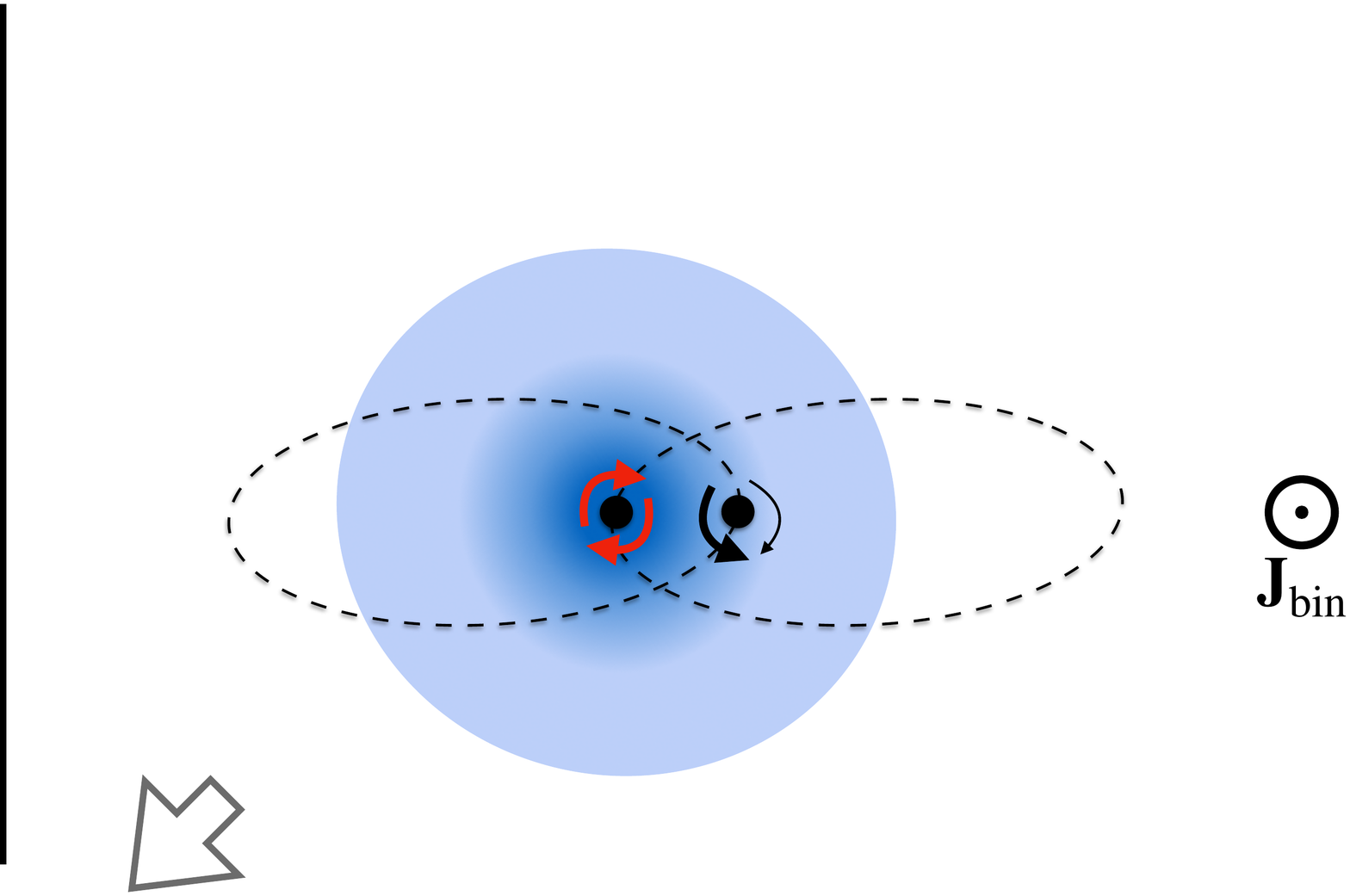}
		\end{minipage}
		
		\vspace{-0.5cm}	
		
		\begin{minipage}[c]{0.5\textwidth}
			\includegraphics[width=\textwidth]{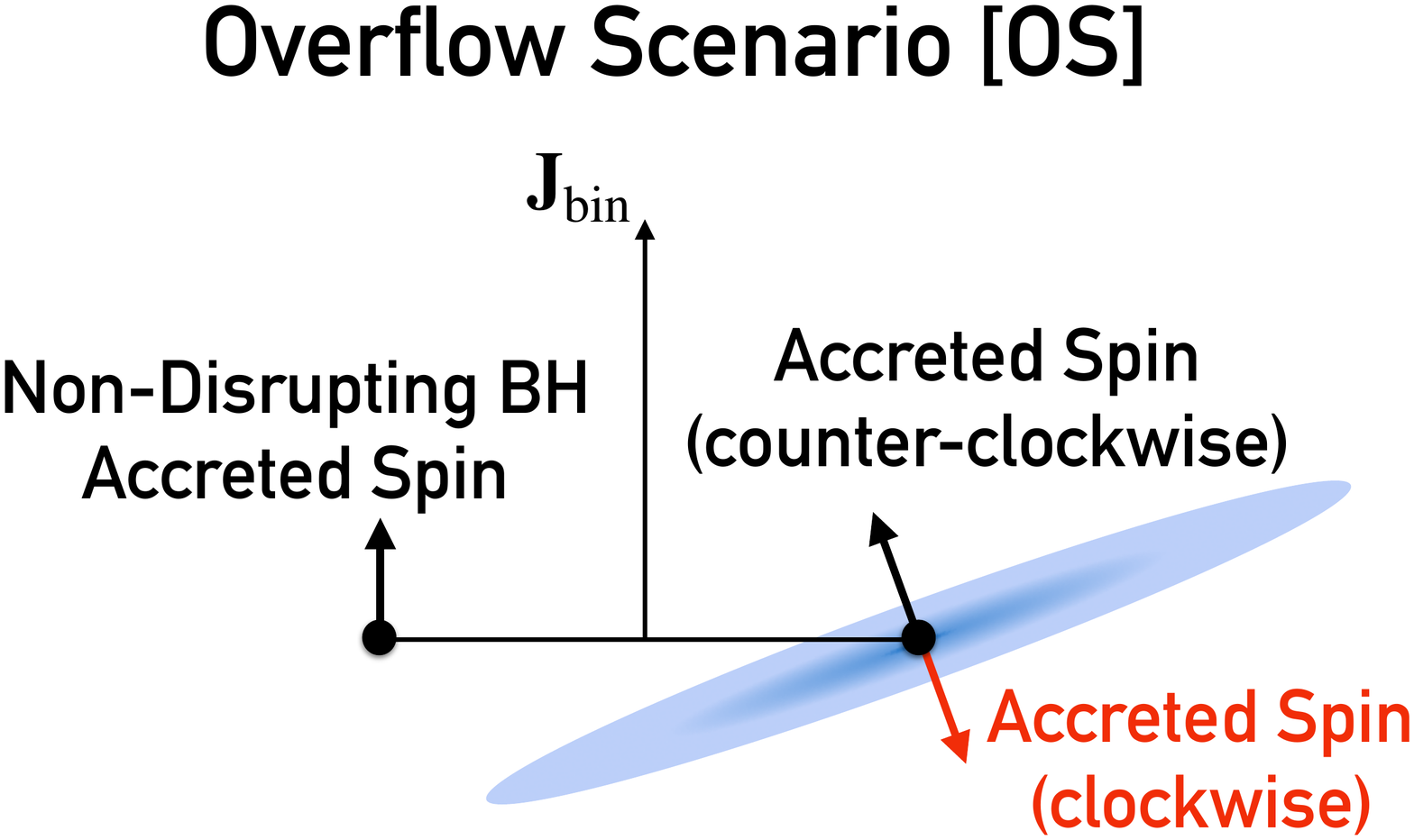}
		\end{minipage}
		
		\vspace{-1.5cm}	
		
	\end{center}
	
	\caption{Two cases are depicted that produce aligned BH spins as well as two cases which give rise to anti-aligned BH spins, all of them belonging to the OS. In all cases, the non-disrupting BH accretes spin in the direction of $J_{\rm bin}$, due to the density gradient it encounters when it enters the accretion disk. The vectors  in the left and right panels indicate the velocity while the vectors in the bottom panels indicate  the spin angular momentum unless noted otherwise. {\it Left and Right Panels:} A view of the LBBH orbital plane before and after a star is disrupted. In each side panels two cases are depicted for the star's orbital motion, which determines the orientation of $\jdisk$. {\it Bottom Panel:} A side view of the LBBH with the final accreted spin directions from both the aligned and anti-aligned configurations. The spin of the non-disrupting BH is always oriented in the direction of $\jbin$. The alignment or anti-alignment of the BH spins is thus mainly determined by the motion of the star before it gets disrupted.}
	\label{fig:OSspins}
\end{figure*}

\subsubsection{Individual Disruptions}
Section \ref{subsec:BBHdyn} outlines the possible scenarios for LBBH TDEs, while Section \ref{subsec:results} shows how the spin magnitude and orientation of each scenario change as a result of these interactions. Following the disruption, accretion disks form around either one or both BHs as shown in Figure \ref{fig:disks}. The angular momentum distribution of material is initially defined by the orbit of the star before disruption, yet  the disk orientation  can be tilted  as  the stream is torqued by the binary \citep{2017MNRAS.465.3840C}.  The misalignment between  $\jbin$ and  $\jdisk$  is expected  to induce a precession of the accretion disk itself \citep{2016LNP...905...45N}. The binary should, over longer timescales,  induce a warped configuration in the disk with a magnitude depending on the local viscosity.  If the accretion disks are misaligned 
with respect to the rotation axis of a Kerr BH, it will be also subject to Lense-Thirring precession \citep{1975ApJ...195L..65B}. The reader is reminded here that a particular LBBH experiencing a TDE might not necessarily merge and that these interactions are expected to only temporarily  alter the spin orientation of the binary. While TDE interactions will  undoubtably change the spin magnitude of the the accreting BHs,  subsequent interactions, expected to take place preferentially with other BHs,  will further modify  $\ceff$.

In Section \ref{subsec:results} we discussed how the accreted spin can go along  $\jbin$ or $\jdisk$ depending on the particular scenario.
\begin{itemize}
	
	\item For the SS, the disrupting BH is the only one that accretes significant stellar debris. The accreted spin is observed to be in the direction of $\jdisk$ at approximately $1.75 \ \rmn{rad}$, which is set by the angular momentum of the star at the time of disruption. 
	
	\item For the CS, the accreted spin of both BHs will be aligned with $\jdisk$. At the time of disruption, $\jdisk$ has an angle of about $2.4 \ \rmn{rad}$ with respect to $\jbin$. As the stream of the most bound material returns to pericenter, the binary torques $\jdisk$ to an angle of $\approx 2 \ \rmn{rad}$. The torqued stream  is responsible for supplying the vast majority of the mass to the disk. As can  be  seen in the left panel of Figure \ref{fig:disks}, the initial stream remains in the disruption plane. 
	
	\item For the OS, the accreted spin of the disrupting BH is in the direction of $\jdisk$ at the time of the initial disruption (at $1.58 \ \rmn{rad}$) while the accreted spin of the non-disrupting BH is aligned with $\jbin$ at angle of $0.24 \ \rmn{rad}$.  The first disruption supplies  the disrupting BH with the majority of
	the accreted mass.  The middle panel of Figure \ref{fig:disks} shows the disk formed by the third disruption  (out of a total of four) whose angle of $\jdisk$ is $0.69 \ \rmn{rad}$ with respect to $\jbin$. 
	
	\item Contrary to the OS where a single BH is responsible for multiple disruptions, the MOS has disruptions occurring onto both BHs sequentially. Out of the two total disruptions, the second and final disruption contributes the majority of mass accreted by the disrupting BH such that the accreted spin is aligned with $\jdisk$ at an angle of $2.2 \ \rmn{rad}$ and the non-disrupting BH accretes spin in the direction of $\jbin$ at an angle $0.14 \ \rmn{rad}$.  The right panel of Figure \ref{fig:disks} shows the disk arising from the first disruption at an angle for $\jdisk$ of $0.85 \ \rmn{rad}$ with respect to $\jbin$. 
	
	\item For the OS and MOS, where multiple disruptions are possible, the angle of $\jdisk$ in Figure \ref{fig:disks} are different from the final angular momentum distribution of the disk. This is because the orientation of disk  changes after each disruption  as a result of the chaotic nature of the three-body dynamics. The disruption resulting in the most accretion will nonetheless determine the final orientation of the BH spins.
	
\end{itemize}

In general, for a subset of LBBH TDEs there is a possibility of relative alignment or anti-alignment between the individual BH spins. Alignment or lack thereof is set by the specific conditions of the stellar disruption as well as by the ensuing orbital dynamics of the binary, as shown in Figures \ref{fig:SSandCSspins} and \ref{fig:OSspins}. For the SS, the interaction is similar to a single BH TDE and only the disrupting BH accretes material and will, as a result, be spun up. Therefore, there will be no spin alignment  between the BHs at the end of the TDE. In this case, the the spin direction of the accreting BH will be  aligned with  $\jdisk$ (Figure \ref{fig:SSandCSspins}). For the CS, the accretion disk is expected to form outside of the binary  such that the  spin directions of both accreting BHs will be similar and aligned with $\jdisk$ (Figure \ref{fig:SSandCSspins}). In the OS, accretion onto each BH is more complicated with the possibility of alignment or anti-alignment. In the case of a single passage  disruption, the spin of the non-disrupting BH will increase in the direction of $\jbin$ as material is accreted. This is because  a steep density gradient  is encountered by the BH  when it enters the disk region, as  illustrated  in Figure \ref{fig:OSspins}. 

\begin{figure*}
	\includegraphics[width=\textwidth]{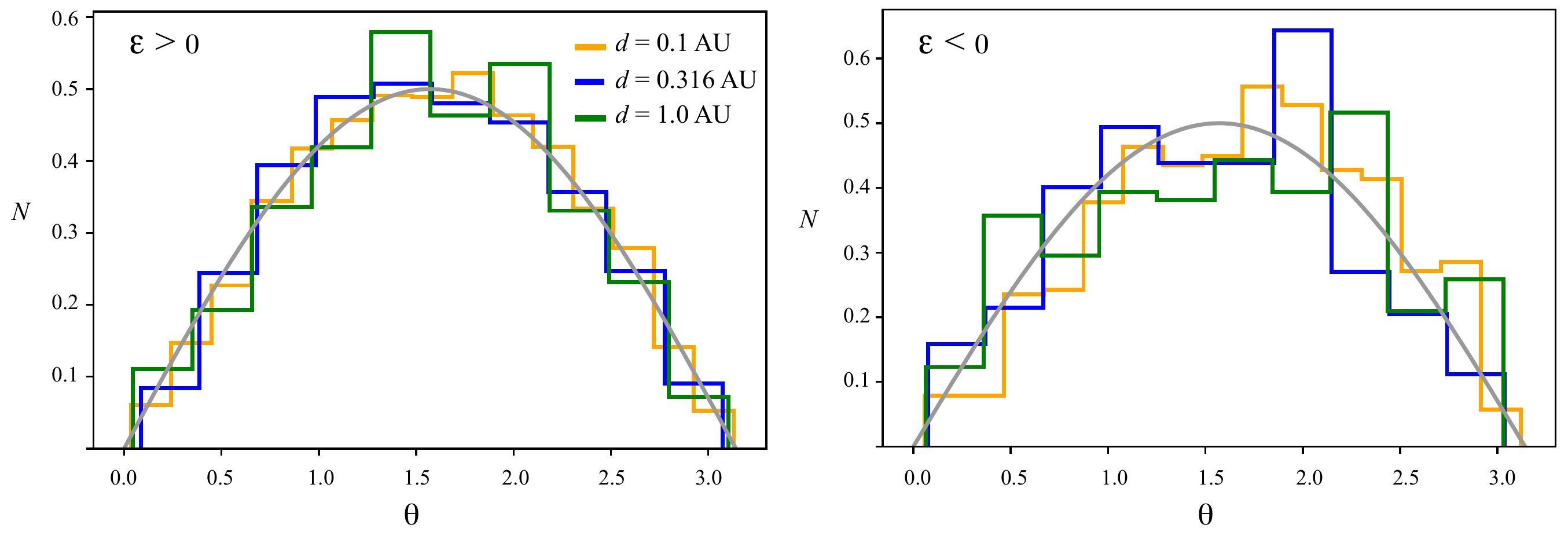}
	\caption{The distributions of relative angles $\theta$ between the stellar velocity vector and the binary orbital velocity vector upon disruption. Similar to Figure~\ref{fig:stats}, the orbital trajectories are calculated for a  sun-like star ($M_\ast=1M_\sun$, $R_\ast=R_\sun$)  interacting with  a $15  \Msun$ equal mass LBBH with $e = 0.5$. The  panels  show the distribution of  $\theta$ for different binary separations ($d=1.0 \rmn{AU} = 87.3 \rtau$, $d=0.316 \rmn{AU} =27.6\rtau$, $d=0.1 \rmn{AU} =8.73 \rtau$). The trajectories for  bound ($\varepsilon<0$) and unbound ($\varepsilon<0$) encounters are plotted separately. For comparison,  an  isotropic   $\theta$ distribution is shown ({\it gray} curve).}
	\label{fig:angledist}
\end{figure*}

The {\it left panels} of Figure \ref{fig:OSspins} shows two cases that produce aligned BH spins:
\begin{itemize}
	\item the star is disrupted outside of the LBBH in the direction of the orbital motion, and  
	\item the star is disrupted inside the LBBH moving against the orbital velocity. 
\end{itemize}
The {\it right panels} of Figure \ref{fig:OSspins} shows two cases that result in anti-alignment:
\begin{itemize}
	\item the star is disrupted outside the LBBH moving against the orbital motion, and 
	\item the star is disrupted inside the LBBH in the direction of the orbital velocity. 
\end{itemize}

We have discussed, in the context of LBBHs, the dynamics and subsequent accretion of stellar debris after a TDE. In all the scenarios, we expect  the direction of the star relative to binary at the moment of disruption  to be an essential parameter in determining the resultant  BH spins. To this end,  we perform a large set of numerical scattering experiments using the $N$-body code developed by \citet{2014ApJ...784...71S} in order to study the distribution of  relative angles between the star's velocity and the binary orbital velocity upon disruption.
The relative angle distributions are plotted in  Figure $\ref{fig:angledist}$ for a sun-like star disrupted by a $15  \Msun$ equal mass BBH  with $e = 0.5$.  From the scattering experiments  we conclude that there is no preferred distribution and, as such, we predict equal probability for alignment and anti-alignment in the OS. It is expected that LBBHs will experience multiple interactions before merging \citep[e.g.,][]{2016PhRvD..93h4029R} and as such, any temporary alignment might be erased before coalescence. TDE interactions from assembly to merge will  nevertheless  alter the spin magnitudes of the the  LBBHs.  It is then tempting to try to constraint the spin properties of LBBHs  experiencing multiple  TDEs and  it is to this issue that we now turn our attention.

\subsubsection{Multiple TDEs and its Relevance to LBBH Growth}
\label{subsubsec:MC}
LIGO has uncovered a population of BHs that is more massive than the population known to reside in accreting binaries \citep{2006ARA&A..44...49R}. One proposed model for the formation of LIGO BHs is through hierarchical mergers of
lighter BHs. In this case, repeated mergers  are expected to  leave a clear imprint on the spin of the final
merger product \citep{2017ApJ...840L..24F, 2017PhRvD..95l4046G, 2018PhRvL.120o1101R, 2019MNRAS.482...30S}. For LBBHs forming hierarchically, the distribution of spin magnitudes is universal and weighted towards high spins. Such a distribution appears to be disfavored  by current observations. This encourages us to investigate spin distributions emerging from LBBHs accreting from multiple TDEs.

\begin{figure}
	\begin{center}
		\includegraphics[width=\linewidth]{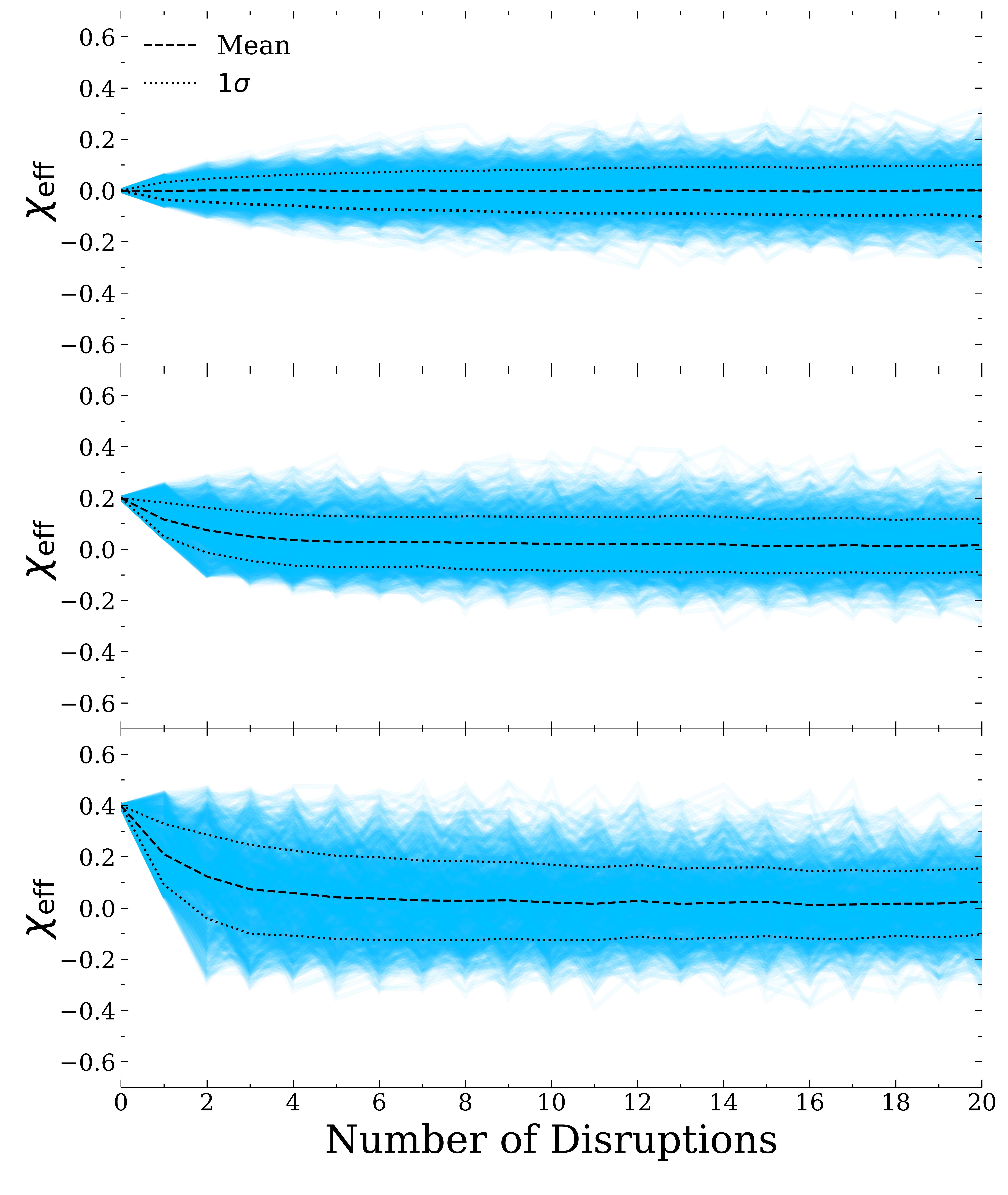}
	\end{center}
	\caption{Three sets of simulations are shown, which are aimed at investigating the evolution of LBBHs that undergo multiple TDEs.  In all cases we plot $\ceff$ as a function of the number of disruptions.  All simulations  start with a LBBH with $\mbh{1} = \mbh{2} = 15 \Msun$ disrupting  $\mstar = 1 \Msun$ stars. The disruptions are assumed to be  isotropically distributed.  For each case,  we change the initial $\ceff$. The {\it top}, {\it middle} and {\it bottom}  panel  start  the binary with $\ceff=0$, $\ceff=0.2$ and $\ceff=0.4$, respectively.}
	\label{fig:MC}
\end{figure}

Three sets of simulations are explored here which are aimed at describing the evolution of LBBHs that undergo multiple TDEs before merging. Each simulation  starts with a binary with $\mbh{1} = \mbh{2} = 15 \Msun$ disrupting  stars with $\mstar = 1 \Msun$ ($q = 0.067$). These binaries  are assumed  to disrupt stars isotropically with respect to $\jbin$.  Then for each set of simulations we change the initial $\ceff$, which is presumed  to be set at BH formation or by the early disruption of a more massive star when the cluster was younger.
Figure \ref{fig:MC} shows our  results. The {\it top panel} initializes the binary with $\ceff=0$, while the {\it middle} and {\it bottom}  panels start the binary  with $\ceff=0.2$ and $\ceff=0.4$, respectively.    For simplicity, we assume the stars are on  parabolic orbits and are fully disrupted in one passage. This results in a total mass accreted of about $0.5 \mstar$ per event, which  is modified  by  an accretion efficiency  that is dependent on the spin of the BH at the time of disruption.
This is done in order to account  for the radiated energy required for a particle at the innermost stable circular orbit to fall into the BH as described in  \citet{1972ApJ...178..347B} and \citet{2017grav.book.....M}. Figure \ref{fig:MC}  shows that if LIGO sources are built up through TDEs, $|\ceff| \lesssim 0.2$ \citep[see also][]{2007arXiv0707.0711M}.  Furthermore, we  show that an  initial  $\ceff$ can be significantly reduced if  BH growth in  the binary  is further promoted  by TDEs.

\subsection{Observable Signatures}
\label{subsec:transients}
A primary source of interest  of TDE interactions has been their prospects  as transients sources. These tidal interactions feed material to the BH at rates that are orders of magnitude  above the  Eddington photon limit (Figure~\ref{fig:OSMOSDIFF}). The total energy, however, is similar from that of other phenomena encountered in astrophysics, and is in fact reminiscent of that released in gamma-ray bursts \citep[GRBs;][]{2009ARAA..47..567G} and canonical TDE jets \citep[e.g.,][]{2012ApJ...760..103D}. One attractive energy extraction mechanism in these systems, which helps  circumvent the Eddington restriction, is the launching of a relativistic jet \citep{2009ApJ...697L..77R,2011MNRAS.416.2102G}. Such flows are able  to carry both bulk kinetic energy and ordered Poynting flux, which  allows high energy radiation to be produced at large distances from the source, where the flow is optically thin \citep[e.g.,][]{2014ApJ...794....9M}.
The corresponding beamed emission offers a  promising observational signature of LBBHs  due to its expected high luminosity.  

\begin{figure}[t]
	\begin{center}
		\includegraphics[width=0.45\textwidth]{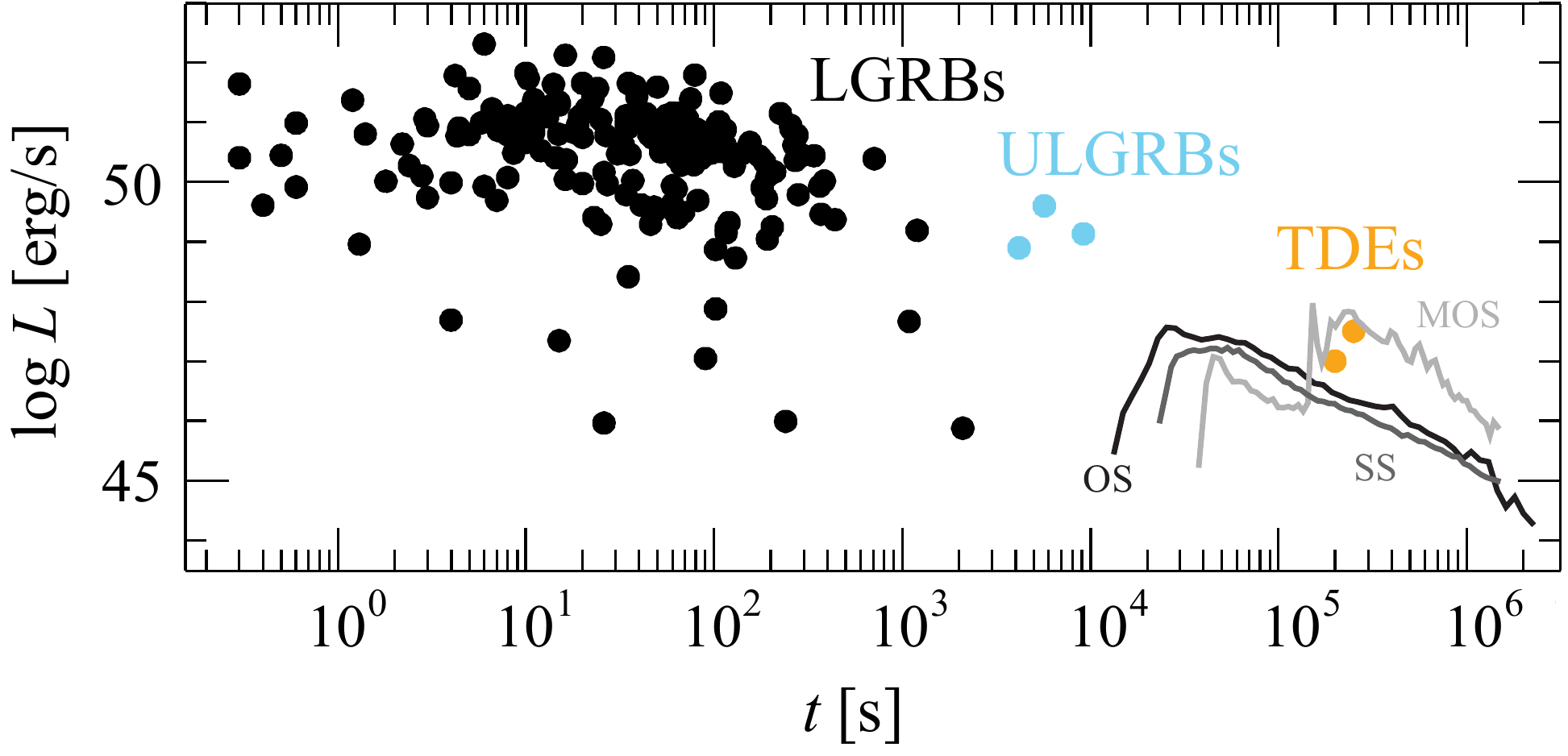}
		\caption{The luminosity  and duration of high energy transients, adapted from \citet{2014ApJ...781...13L}. Shown are the predicted luminosities of  three of the scenarios for LBBH TDEs
			discussed in this paper, assuming $L_{\rm j}\propto \dot{\rmn{M}}$ and a 10\% radiative efficiency. For comparison we plot  the observed high-energy properties of GRBs and jetted TDEs.  The timescales and durations of LBBH TDEs are well removed from typical long GRBs, but lie  between  those of the emerging class of ultra-long GRBs and jetted TDEs.}
		\label{fig:phasespace}
	\end{center}
	\vspace{-0.3in}
\end{figure}

Figure \ref{fig:phasespace} shows the predicted  luminosities for stars disrupted by LBBHs assuming that the jet power  traces the mass supply to the BH: $L_{\rm j} \propto \dot{\rmn{M}} c^2$.  For comparison we also plot  the luminosities and durations of long $\gamma$-ray bursts (LGRBS), jetted TDEs from galactic nuclei  as well as those from the newly emerging class of ultra-long GRBs: GRB 101225A, GRB 111209A, and GRB 121027A \citep{2014ApJ...781...13L}. 
These ultra-long GRBs reach peak X-ray luminosities of $\approx 10^{49} \text{erg s}^{-1}$ and show non-thermal spectra that is  reminiscent of relativistically beamed emission. The derived properties of these LBBH TDEs
appear to place them between ultra-long GRBs and  jetted TDEs from galactic nuclei. Our ability to classify long duration transients as events emanating from LBBHs or massive BHs  in galactic nuclei  is likely to  remain a challenge. One alternative  in the near term is to search  at the astrometric positions of these long transients and see whether they  are coincide with galactic centers.

Another idea is to look for interruptions in the observed light curve caused by the binary companion, from which one could extract the orbital time of the disrupting BBH and thereby its orbital parameters \citep[e.g.,][]{2014ApJ...786..103L}. The relativistically beamed emission from these events is the only component that might be readily detectable since the disk emission is expected to be Eddington limited. We therefore conclude  that one avenue for constraining  whether or not LBBHs  reside  in star clusters is searching for their high-energy signatures. The possibility of collecting a sample of such events in coming years with {\it Swift} appears promising, provided that the rate is similar to the LIGO merger rate of LBBHs \citep[for a detailed discussion on detectability the reader is refer to][]{2014ApJ...794....9M}.

To get an estimate on the LBBH TDE rate from the GC population we start by computing the rate per
GC using $\Gamma_{\rm TDE} \approx N_{\rm BBH} \times \eta_{\rm s} \sigma_{\rm TDE} v_{\rm dis}$, where $N_{\rm BBH}$ is the number of BBHs per GC, $\eta_{\rm s}$ is the number density of single stars,  $\sigma_{\rm TDE}$ is the TDE cross section, and $v_{\rm dis}$ is the cluster velocity dispersion. The cross section $\sigma_{\rm TDE}$ can be written as a product of the binary-single interaction cross section and the probability for an
interaction to result in a TDE \citep[e.g.][]{2017ApJ...846...36S},
i.e. $\sigma_{\rm TDE} \approx \sigma_{\rm bs} \times P_{\rm TDE}$. Assuming the
gravitational focusing limit for $\sigma_{\rm bs}$ and $P_{\rm TDE} \approx 2 \rtau/a$ one finds,
\begin{equation*}
\Gamma_{\rm TDE}^{\rm gal.} \approx 10^{-6}\ \text{yr}^{-1}\ \left(\frac{\eta_{\rm s}}{10^{5}\text{pc}^{-3}}\right) \left(\frac{\Mbh}{30\Msun}\right)^{4/3}\left(\frac{15\text{km/s}}{v_{\rm dis}}\right),
\end{equation*}
where this rate is per  galaxy ($5$ LBBHs per GC, and $200$ GCs per galaxy) derived
for solar type stars ($1\Msun,1\Rsun$)
interacting with LBBHs of equal mass. This is about one order of magnitude smaller than the LIGO merger rate, and
will be further observationally suppressed due to the expected beaming. Therefore, we expect observations of beamed LBBH TDEs to be  lower than the inferred LBBH merger rate. However, if one instead considers stellar tidal disruptions by single
BHs in GCs the rate of beamed TDEs is higher by a factor roughly given by the number ratio of single BHs to the number of LBBHs,
\begin{equation*}
\Gamma_{\rm TDE}^{\rm BH} \approx \Gamma_{\rm TDE}^{\rm LBBH} \times \frac{N_{\rm BH}}{N_{\rm BBH}},
\end{equation*}
where $\Gamma_{\rm TDE}^{\rm BH}$ ($\Gamma_{\rm TDE}^{\rm LBBH}$) is the rate from single (binary) BH stellar disruptions.
Assuming the fraction of LBBHs to be at the percent level then this leads to that the rate of stellar single BH TDEs is
$\approx 10^{-4}\ \text{yr}^{-1}$ per galaxy, which is much closer to observable limits. This scenario
was recently studied in \cite{2016ApJ...823..113P},
and might also be used to constrain the BH population that later forms LBBHs. We note that our estimate might be at the
optimistic side compared to the rates derived in \cite{2016ApJ...823..113P}, but any of these
estimates should be taken with caution and more sophisticated $N$-body methods must be used to  explore this further. 

Irrespective of current uncertainties, the detection or non-detection of long duration transients from BH and LBBH stellar disruptions should  offer strong constraints on the population of LBBHs and the nature of the stellar clusters that host them. In an upcoming paper we explore what the characteristic LBBH orbital parameters are for different cluster types, as well as what we can learn about the dynamical formation of LBBH GW sources from observing the associated population of BH and LBBH TDEs.

\section*{Acknowledgments}
\label{sec:thanks}

The authors thank  S. Schr{\o}der, T. Fragos, B. Mockler, S.  I. Mandel, W. Farr, C. Miller, D. J. D'Orazio, K. Hotokezaka, M. Gaspari and A. Askar  for stimulating discussions.
MLJR acknowledges that all praise and thanks belongs to Allah (any benefit is due to God and any shortcomings are my own).
ERR acknowledge support from the DNRF (Niels Bohr Professor)  and  NSF  grant  AST-1615881.
JS acknowledges support from the Lyman Spitzer Fellowship.
The authors further thanks the Niels Bohr Institute for its hospitality while part of this work was
completed, and the Kavli Foundation and the DNRF for supporting the 2017 Kavli Summer Program.

\bibliography{References.bib}

\end{document}